\documentclass[11pt,amssymb,amsfonts,amscd]{amsart}

\usepackage{amscd}
\usepackage{amssymb}
\usepackage{amsmath}
\newtheorem{proposition}{Proposition}[section]
\newtheorem{lemma}[proposition]{Lemma}
\newtheorem{definition}[proposition]{Definition}
\newtheorem{theorem}[proposition]{Theorem}
\newtheorem{cor}[proposition]{Corollary}

\newtheorem{example}[proposition]{Example}

\newcounter{THNO}[section]
\newcounter{SNO}[section]
\newcounter{multieq}[equation]

\newcounter{tmp}

\def\h#1,#2{{\rm Hom}({#1}\:,\; {#2})}
\def\H#1,#2,#3,#4{{\rm Hom}^{#1}_{#2}({#3}\:,\; {#4})}
\def\E#1,#2,#3,#4{{\rm Ext}^{#1}_{#2}({#3}\:,\; {#4})}
\def\lto{\longrightarrow}

\def\da{\big\downarrow}
\def\ua{\big\uparrow}

\def\ss#1{\scriptsize{$#1$}}

\def\ot{\otimes}

\def\op{\oplus}
\def\M{{\mathfrak M}}
\def\wh#1{\widehat{#1}}
\def\wt#1{\widetilde{#1}}
\def\wb#1{\bar{#1}}
\def\R{{\mathbb{R}}}
\def\C{{\mathbb{C}}}
\def\Q{{\mathbb{Q}}}
\def\Z{{\mathbb{Z}}}
\def\T{{{\rm T}}}
\def\P{{\mathbb{P}}}

\def\llongrightarrow{{\;\relbar\joinrel\relbar\joinrel\relbar\joinrel
\rightarrow\;}}

\def\O{{\mathcal{O}}}
\def\E{{\mathcal{E}}}
\def\F{{\mathcal{F}}}
\def\G{{\mathcal{G}}}
\def\H{{\mathcal{H}}}

\def\J{{\mathcal{J}}}

\def\End{{\rm E}{\rm n}{\rm d}}
\def\Hom{{\rm H}{\rm o}{\rm m}}
\def\Aut{{\rm A}{\rm u}{\rm t}}
\def\hom{\underline{{\mathcal H}om}}
\def\ext{\underline{{\mathcal E}xt}}
\def\GL{{\rm G}{\rm L}}
\def\SO{{\rm S}{\rm O}}

\def\Ext{{\rm E}{\rm x}{\rm t}}
\def\RHom{{\bf R}{\rm H}{\rm o}{\rm m}}
\def\rhom{{\bf R}\underline{{\mathcal H}om}}
\def\h{{\hbar}}

\def\Ko{{\mathcal C}^{\cdot}}

\def\Mu{{\mathit J}}
\def\Nu{{\mathit I}}
\def\mui{{\mathrm j}}
\def\nui{{\mathrm i}}

\def\CK{{\mathcal K}}

\def\Om{{\varOmega}}
\def\PTh{{\mathbb{P}}^{3}_{\h}}
\def\PT{{\mathbb{P}}^{3}}
\def\Pl{{\mathbb{P}}^{2}}
\def\pth{{PS_{\h}}}
\def\Plh{{\mathbb{P}}^{2}_{\h}}
\def\plh{{PP_{\h}}}
\def\S{{\mathbb{S}}}

\def\be{\begin{equation}}
\def\ee{\end{equation}}

\newcommand{\CC}{{\mathbb C}}
\newcommand{\RR}{{\mathbb R}}
\newcommand{\ZZ}{{\mathbb Z}}
\newcommand{\PP}{{\mathbb P}}
\newcommand{\Tr}{{\rm Tr}}
\newcommand{\diag}{{\rm diag}}
\newcommand{\ra}{\rightarrow}

\newcommand{\cA}{{\mathcal A}}

\newcommand{\twice}{{\cdot\cdot}}

\newcommand{\Coker}{\mathop{\mathbf{Coker}}\nolimits}
\newcommand{\Image}{\mathop{\mathbf{Im}}\nolimits}

\newcommand{\id}{\mathbf{id}}

\newcommand{\Ker}{\mathop{\mathbf{Ker}}\nolimits}
\renewcommand{\Im}{\mathop{\mathbf{Im}}\nolimits}
\newcommand{\gr}{\mathcal{G}}
\newcommand{\fl}{\mathcal{FL}}
\newcommand{\Gr}{\mathbf{G}}
\newcommand{\Fl}{\mathbf{Fl}}

\newcommand{\cz}{{\check z}}

\newcommand{\CV}{{\mathcal{V}}}
\newcommand{\CO}{{\mathcal{O}}}
\newcommand{\FS}{{\mathfrak{S}}}
\newcommand{\CS}{{\mathcal{S}}}
\newcommand{\otR}{{\ \underset{R}{\ot}\ }}
\newcommand{\timesR}{{\ \underset{R}{\times}\ }}
\newcommand{\New}{{\mathcal N}}
\newcommand{\dia}{{\diamond}}

\def\Vec{{U}}
\def\kk{{\mathtt k}}
\def\ctwo{{k}}
\def\bnul{{b_{1}}}
\def\bone{{b_{2}}}
\def\Bnul{{B_{1}}}
\def\Bone{{B_{2}}}

\newcommand{\TT}{{\mathcal T}}
\newcommand{\Vect}{{\mathcal Vect}}
\newcommand{\Comm}{{\mathcal Comm}}

\newcommand{\bz}{{\bar z}}
\newcommand{\prh}{{\widehat{\pth}}}
\newcommand{\bxi}{{\bar\xi}}

\newcommand{\Spr}{{S'}}
\newcommand{\Ssta}{{S^{\vee}}}

\begin{document}

\begin{titlepage}

\renewcommand{\thepage}{ }
\renewcommand{\today}{ }

\thispagestyle{empty}
\vspace*{-10pt}
{\small \hfill{IASSNS-HEP-00/14}}

\vspace{1.4in}

\title{Noncommutative instantons and twistor transform}

\author{Anton Kapustin}
\thanks{The first author was supported by DOE grant DE-FG02-90ER4054442.}
\address{
 School of Natural Sciences, Institute for Advanced Study,
Olden Lane, Princeton, NJ 08540
}
\email{kapustin@ias.edu}

\author{Alexander Kuznetsov}
\thanks{The second author
was  supported by NSF grant DMS97-29992 and RFFI grants 99-01-01144,
99-01-01204.}
\address{
Institute for Problems of Information Transmission, 
 Russian Academy of Sciences,
19 Bolshoi Karetnyi, Moscow 101447, Russia
}
\email{sasha@kuznetsov.mccme.ru, akuznet@ias.edu}

\author{Dmitri Orlov}
\thanks{The third author was supported by NSF grant DMS97-29992 and RFFI grant
99-01-01144.}
\address{
Algebra Section, Steklov Mathematical Institute, 
Russian Academy of Sciences, 
 8 Gubkin str., GSP-1, Moscow 117966, Russia
}
\email{orlov@mi.ras.ru}

\date{24 February, 2000}

\dedicatory{Dedicated to A.N. Tyurin on his 60th birthday} 
\maketitle 

\begin{sloppypar}
\vspace{50pt} 
\centerline{\large\bf Abstract}
\vspace{15pt}

\noindent 
Recently N.~Nekrasov and A.~Schwarz proposed a modification of the ADHM construction of 
instantons which produces instantons on a noncommutative deformation of $\RR^4$. 
In this paper
we study the relation between their construction
and algebraic bundles on noncommutative projective spaces. We exhibit 
one-to-one correspondences between three classes of objects: framed bundles on a  
noncommutative $\PP^2,$ certain complexes of sheaves on a noncommutative $\PP^3,$ and 
the modified ADHM data. The modified ADHM construction itself is interpreted in terms of 
a noncommutative version of the twistor transform. We also prove that the moduli space of 
framed bundles on the noncommutative $\PP^2$ has a natural hyperk\"ahler metric and is
isomorphic as a hyperk\"ahler manifold to the moduli space of framed torsion free sheaves on 
the commutative $\PP^2.$ The natural complex structures on the two moduli spaces do not
coincide but are related by an $\SO(3)$ rotation. Finally, we propose a construction
of instantons on a more general noncommutative $\RR^4$ than the one considered
by Nekrasov and Schwarz (a $q$\!--\,deformed $\RR^4$).


\end{sloppypar}

\end{titlepage}

\section{Physical motivation}

In this section we explain the physical motivation for studying instantons
on a noncommutative $\RR^4$. 
Readers uninterested in the motivation may skip most of this section
and proceed directly to subsection~\ref{noncomminst}. Likewise, readers
familiar with the way noncommutative instantons arise in string theory may
start with subsection~\ref{noncomminst}.

\subsection{Instanton equations}

Let $E$ be a vector bundle with structure group $G$ on an oriented
Riemannian 4-manifold $X,$ and let $A$ be a connection on $E$. Instanton equation is
the equation
\begin{equation}\label{inst}
F_A^+=0,
\end{equation} 
where $F_A$ is the curvature of $A,$ and $F_A^+$ denotes the self-dual (SD) part of $F_A$. 
Solutions of this equation are called instantons, or anti-self-dual (ASD) connections. 
The second Chern class of $E$ is known in the physics literature as the instanton number. 
Instantons automatically
satisfy the Yang-Mills equation $d_A(*F)=0,$ where $d_A: \Omega^{p}\ot
\End (E)\lto \Omega^{p+1}\ot \End (E)$ is the covariant differential, and
$*:\Omega^p\lto \Omega^{4-p}$ is the Hodge star operator.

There are several physical reasons to be interested in instantons. If one
is studying quantum gauge theory on a Riemannian 3-manifold $M$
(space), then instantons on $X=M\times \RR$ describe quantum-mechanical
tunneling between different classical vacua. The possibility of such tunneling
has drastic physical effects, some of which can be experimentally 
observed. If one is studying classical gauge theory on a 5-dimensional space-time
$X\times \RR,$ then instantons on $X$ can
be interpreted as solitons, i.e. as static solutions of the Yang-Mills
equations of motion. In fact, instantons are the absolute minima of the
Yang-Mills energy function of the 5-dimensional theory (with fixed
second Chern class). 

Both interpretations arise in string theory, but to explain this we need to
make a digression and discuss D-branes. 

\subsection{D-branes}

It has been discovered in the
last few years that string theory describes, besides strings, 
extended objects (branes) of various dimensions. These extended objects should
be regarded as static solutions of (as yet poorly understood) stringy equations of
motion. D-branes are a particularly manageable class of branes. Recall that
ordinary closed oriented superstrings, known as Type II strings, are described 
by maps from a Riemann surface $\Sigma$ (``worldsheet'') to a 10-dimensional manifold $Z$
(``target''). The physical definition of a D-brane is ``a submanifold of $Z$ on 
which strings can end.'' This means that if a D-brane is present, then one needs to 
consider maps from a Riemann surface with boundaries to $Z$ such that the 
boundaries are mapped to a certain submanifold $X\subset Z$. In this case one 
says that there is a D-brane wrapped on $X$. If $X$ is connected and has 
dimension $p+1,$ then one says that one is dealing with a Dp-brane. In general, 
$X$ can have several components with different dimensions, and then each 
component corresponds to a D-brane.

In perturbative string theory, the role of equations of motion is played
by the condition that a certain auxiliary quantum field theory on
the Riemann surface $\Sigma$ is conformally invariant. When D-branes are 
present, $\Sigma$ has boundaries, and the auxiliary theory must be 
supplemented with boundary conditions. The requirement that the boundary
conditions preserve conformal invariance imposes constraints on the
submanifold $X$. This constraints should be regarded as equations of
motion for D-branes. For example, if we consider a D0-brane wrapped on
a 1-dimensional submanifold $X,$ then conformal invariance requires
that $X$ be a geodesic in $Z$. This is the usual equation of motion for
a relativistic particle moving in $Z$.

An important subtlety is that to specify fully the boundary conditions for the auxiliary 
theory on $\Sigma$ it is not sufficient to specify $X;$ one should
also specify a unitary vector bundle $E$ on $X$ and a connection on it. In the 
simplest case this bundle has rank $1,$ but one can also have ``multiple'' D-branes, described by 
bundles of rank $r>1$. Such bundles describe $r$ coincident D-branes wrapped 
on the same submanifold $X$. Using the requirement of conformal invariance of the auxiliary
two-dimensional quantum field theory, one can derive equations of motion for
the Yang-Mills connection on $E$. In the low-energy approximation, the equations of motion 
are the usual Yang-Mills equations $d_A(*F_A)=0.$ In particular, instantons are solutions of
these equations.

\subsection{Instantons and D-branes}\label{instanddbranes}

Let $Z$ be $\RR^{10}$ with a flat metric, and let $X\hookrightarrow Z$ be 
$\RR^5=\RR^4\times \RR$ linearly embedded in $Z$. We regard $\RR^4$ as space and $\RR$ as time.
Consider $r$ D4-branes wrapped on $X$. This physical system is described
by the Yang-Mills action on $\RR^5=\RR^4\times \RR$. If one is looking
for static solutions of the equations of motion, one needs to consider
the minima of the Yang-Mills energy function
$$
W[A]=\int_{\RR^4}\ ||F_A||^2,
$$
where $F_A$ is the curvature of a $U(r)$ connection $A,$ and
$||F_A||^2=-\Tr\left(F_A\wedge *F_A\right).$ The instanton number of
$A$ is defined by
\begin{equation}\label{instnumber}
c_2= \frac{1}{8\pi^2}\int_{\RR^4}\ \Tr\left(F_A\wedge F_A\right).
\end{equation}
If the Yang-Mills energy evaluated on $A$ is finite, then the bundle $E$
and the connection $A$ extend to $\S^4$, the one-point compactification of
$\RR^4$ (see~\cite{Atiyah} for details). In this case $c_2$ is the second Chern 
class of $E$ and is therefore an integer.

Solutions of instanton equations on $\RR^4$ are precisely the absolute minima
of the Yang-Mills energy function.
These solutions should be regarded as composed  of identical particle-like 
objects (instantons) on $X,$ their number being $c_2$. Since
the energy of the instanton is proportional to $c_2,$ all ``particles''
have the same mass. Since the solution is static, the particles neither repel
nor attract. This is actually a consequence of supersymmetry: Type II
string theory is supersymmetric, and D4-branes with instantons on them leave
part of supersymmetry unbroken.

In string theory one may also consider $k$ D0-branes present simultaneously
with $r$ D4-branes. More specifically, we will consider D0-branes which are at
rest, i.e.  the corresponding one-dimensional manifolds are straight lines
parallel to the time axis. Such a configuration of branes is also
supersymmetric, and consequently there are no forces between any of the branes.
The positions of D0-branes are not constrained by anything, so their 
moduli space is $(\RR^9)^k$. More precisely, since D0-branes are 
indistinguishable, the moduli space is $Sym^k(\RR^9)$. 

It turns out that an instanton with instanton number $k$ and $k$ D0-branes
are related: they can be deformed into each other without any cost in energy. 
A convenient point of view is the following.
In the presence of D4-branes wrapped on $X$ the moduli space of D0-branes has two branches: a 
branch where their positions
are unconstrained and D0-branes are point-like (this branch is isomorphic to 
$Sym^k(\RR^9)$), 
and the branch where they are constrained to lie on $X$. The latter branch is 
isomorphic to the moduli space ${\mathcal M}_{r,k}$ of $U(r)$ instantons on $X=\RR^4$ with $c_2=k$. 

The dimension of ${\mathcal M}_{r,k}$ is known to be $4rk$ for $r>1$~(see 
for example~\cite{Atiyah}).
For $r=1$ instantons do not exist. The translation group of $\RR^4$ acts freely on
${\mathcal M}_{r,k},$ and the quotient space describes the relative positions and sizes of
instantons. Thus D0-branes are point-like objects when they are away from D4-branes, 
but when they bind to D4-branes they can acquire finite size. 

The ``instanton'' branch touches the ``point-like'' branch at submanifolds
where some or all of the instantons shrink to zero size. These are the submanifolds
where the instanton moduli space is singular. At these submanifolds
the point-like instantons can detach from D4-branes and start a new life as D0-branes. 
This lowers the second Chern class of the bundle on D4-branes. Thus
from the string theory perspective it is natural to glue together the moduli
spaces of instantons with different Chern classes along singular submanifolds.

\subsection{Noncommutative geometry and D-branes}

Instanton equations (and, more generally, Yang-Mills equations) arise in the low-energy 
limit of string theory, or equivalently for large string tension. Recently, another kind 
of low-energy limit of string
theory was discussed in the literature~\cite{SeibergWitten}. Consider
a trivial $U(r)$-bundle on $X=\RR^4$ with a connection $A$ whose curvature $F_A$
is of the form $1\ot f,$, where $1$ is the unit section of $\End (E)$, and
$f$ is a constant nondegenerate 2-form. For small $f$ 
the D4-branes are described by the ordinary Yang-Mills action, but for 
large $F_A$ the stringy equations of motion get complicated. 
It turns out that the equations of motion simplify again in the limit when 
both $F_A$ and the string tension are taken to infinity, with a certain 
combination of the two kept fixed (one also has to scale the metric
appropriately, see \cite{SeibergWitten}). We will call this limit the Seiberg-Witten limit. 
In this limit the D4-branes are described 
by Yang-Mills equations on a certain noncommutative deformation of $\RR^4$ 
(see \cite{SeibergWitten} and references therein).

There is another description of the Seiberg-Witten limit, which is gauge-equivalent to the
previous one. Type II string theory reduces at low energies to Type II supergravity
in 10 dimensions. The bosonic fields of this low-energy theory include
a symmetric rank-two tensor (metric) and a 2-form $B$. $\RR^{10}$ with a
flat Lorenzian metric and a constant $B$ is a solution of supergravity equations
of motion, as well as full stringy equations of motion. A constant $B$ can be gauged
away, so this is not a very interesting solution. Life gets more interesting
if there are D-branes present. For example, consider $r$ coincident flat D4-branes 
embedded in $\RR^{10}$ with a constant $B$-field. It turns out that one can gauge away a 
constant $B$-field only at the expense of introducing a constant $F_A$ of the form
$1\ot f$ where $f$ is equal to the pull-back
of $B$ to the worldvolume of the D4-branes. Thus the solution with zero $F_A$ and nonzero 
$B$ is equivalent to the solution with nonzero $F_A$ and zero $B$. Therefore the 
Seiberg-Witten limit can be described as the limit in which both the $B$-field and the 
string tension become infinite. 

The idea that D-branes in a nonzero B-field are described Yang-Mills theory on a 
noncommutative space was first put forward in~\cite{CDS} for the case of D-branes
wrapped on tori. 

\subsection{Instanton equations on a noncommutative $\RR^4$}\label{noncomminst}

The deformed $\RR^4$ that one obtains in the Seiberg-Witten limit is completely 
characterized by its
algebra of functions $\cA$. It is a noncommutative algebra whose underlying
space is a certain subspace of $C^\infty$ functions on $\RR^4$. The product
is the so-called Wigner-Moyal product formally given by
\begin{equation}\label{wm}
(f\star g)(x)=\lim_{y\ra x} 
\exp\left(\frac{1}{2}\h\theta_{ij}\frac{\partial^2}{\partial x_i\partial y_j}\right)
f(x)g(y).
\end{equation}
Here $\theta$ is a purely imaginary matrix, and $\h$ is a 
real parameter (``Planck constant'') which is introduced to emphasize that the 
Wigner-Moyal product is a deformation of the usual product. In the string theory
context $\theta$ is proportional to $f^{-1}.$

Of course, to make sense of this 
definition we must specify a subspace in the space of $C^\infty$ functions which is closed under 
the Wigner-Moyal product. Leaving this question aside for a moment,\footnote{String theory 
considerations do not shed light on this problem.} one can define the
exterior differential calculus over $\cA$. Differential geometry of noncommutative spaces
has been developed by A.~Connes~\cite{Connes}. In our situation Connes' general
theory is greatly simplified. For example, the sheaf of 1-forms
$\Omega^1(\cA)$ is simply a bimodule $\cA^{\oplus 4}$ (the relation of this definition
with the general theory is explained in subsection~\ref{differentials}). 
The elements of $\Omega^1(\cA)$ will be 
denoted $\sum_i f^i(x) dx_i,$ or simply $f^i(x) dx_i$. The exterior differential $d$ is a 
vector space morphism
$$
d: \cA\ra \Omega^1(\cA),\qquad f \mapsto \frac{\partial f}{\partial x_i} dx_i.
$$
The exterior differential $d$ satisfies the Leibniz rule
$$
d(f_1\star f_2)=df_1\star f_2+ f_1\star df_2.
$$
This makes sense because $\Omega^1(\cA)$ is a bimodule.

The sheaf of 2-forms over $\cA$ is a bimodule $\Omega^2(\cA)=\cA^{\oplus 6}$
(see subsection~\ref{differentials}). 
The definition of the exterior differential can be extended to $\Omega^1(A)$ in an 
obvious manner.

Complex conjugation acts as an anti-linear anti-homomorphism of $\cA$, i.e.
$\overline{(f\star g)}=\overline{g}\star \overline{f}.$ Thus $\cA$ has a natural structure 
of a $*$-algebra. We will denote the $*$-conjugate of $f\in\cA$ by $f^\dagger$.

A trivial bundle over the noncommutative $\RR^4$ is defined as a free $\cA$-module $E$.
A trivial unitary bundle over the noncommutative $\RR^4$ is defined as 
a free module $V\ot_\CC \cA,$ where $V$ is a Hermitean vector space.
A connection on a trivial bundle $E$ is defined as a map
$$
\nabla:\ E\ra E\ot_\cA \Omega^1(\cA),
$$
which is a vector space morphism satisfying the Leibniz rule
$$
\nabla(m\star f)=\nabla(m)\star f+m \star df.
$$
This formula makes use of the bimodule structure on $\Omega^1(\cA)$.

The curvature $F_\nabla=[\nabla,\nabla]$ is a morphism of $\cA$-modules 
$$
F_\nabla: E\ra E\ot_\cA \Omega^2(\cA).
$$
As in the commutative case, a connection on a trivial bundle $E$ can be written in 
terms of a connection 1-form $A\in \End_{\cA}(E)\ot_\cA\Omega^1(\cA)$:
$$
\nabla(m)=dm+A\star m.
$$
This formula uses the bimodule structure on $m$.
If $E$ is a unitary bundle, and we have $A^\dagger=-A$, then we say that $A$ is a unitary 
connection.

The curvature is given in terms of $A$ by the usual formula
$$
F_\nabla:= F_A=d A+A\wedge A.
$$
Here it is understood that
$$
f^i\ dx_i\ \wedge\ g^j dx_j=f^i\star g^j\ dx_i\wedge dx_j.
$$
The instanton equation on $\cA$ is again given by~(\ref{inst}), and the instanton number is
defined by~(\ref{instnumber}).

The most obvious choice of the space of functions closed under the Wigner-Moyal
product is the space of polynomial functions. However, this choice
is not suitable for our purposes because it precludes the decrease of $F_A$
at infinity which is necessary for the instanton action to converge.
In the commutative case, components of an instanton connection are rational 
functions~\cite{Atiyah}, so we would like our class of functions to include 
rational functions on $\RR^4$. A possible choice for the underlying set of $\cA$ is the set of 
$C^\infty$ functions on $\RR^4$ all of whose derivatives are polynomially bounded. 
Then we face the question of
the convergence of the series~(\ref{wm}). To avoid dealing with this issue, we modify
our definition of the Wigner-Moyal product (see Appendix for details). The modified
product makes the space of $C^\infty$ functions all of whose derivatives are 
polynomially bounded into an algebra over $\CC,$ and agrees with~(\ref{wm}) 
on polynomial functions.

Polynomial functions form a subalgebra of $\cA$.
This subalgebra is isomorphic to the algebra generated by four variables 
$x_i,i=1,2,3,4$ with relations
$$
[x_i,x_j]=\h\theta_{ij}.
$$
This algebra is usually called the Weyl algebra.

To summarize, there is a limit of string theory in which D4 branes are described
by Yang-Mills equations on the noncommutative $\RR^4$ ($=\cA$).
D0-branes bound to D4-branes are described in this limit by the instanton
equations on the noncommutative $\RR^4$. One can show that, unlike 
in the commutative case, instantons cannot be deformed to point-like
D0-branes without a cost in energy. Thus it is natural to
suspect that the moduli space of instantons on the noncommutative $\RR^4$
is metrically complete.

\section{Review of the ADHM construction and summary}\label{summary}

All instantons on the commutative $\RR^4$ arise from the so-called
ADHM construction. Recently N.~Nekrasov and A.~Schwarz~\cite{NS} introduced a modification
of this construction which produces instantons on the noncommutative 
$\RR^4$.\footnote{As in the commutative case, one may consider different classes
of functions on the noncommutative $\RR^4$: polynomial, $C^\infty$ functions rapidly
decreasing at infinity, $C^\infty$ functions all of whose derivatives are polynomially
bounded, etc. Our class of functions differs somewhat from that adopted
in~\cite{NS}.}
In the commutative case the completeness of the ADHM construction
can proved using the twistor transform of R.~Penrose, so one could hope that that
the same approach could work in the noncommutative case. In this paper we show that the 
deformed ADHM data of~\cite{NS} describe holomorphic bundles on certain noncommutative
algebraic varieties and interpret the deformed ADHM construction in terms of 
noncommutative twistor transform. In this subsection we review both ordinary and deformed
ADHM constructions and make a summary of our results.

\subsection{Review of the ADHM construction of instantons}

First let us outline the ADHM construction of $U(r)$ instantons on the commutative $\RR^4$ 
following~\cite{Do}. We assume that the constant metric $G$ on $\RR^4$ has been brought
to the standard form $G=\diag(1,1,1,1)$ by a linear change of basis.
To construct a $U(r)$ instanton with $c_2=k$ one starts with two
Hermitean vector spaces $V\simeq \CC^k$ and $W\simeq \CC^r$. The ADHM data
consist of four linear maps $B_1,B_2\in \Hom(V,V),\ I\in \Hom(W,V),\ J\in \Hom(V,W)$
which satisfy the following two conditions:

(i) $\mu_c= [B_1,B_2]+IJ=0,\quad \mu_r= [B_1,B_1^\dagger]+[B_2,B_2^\dagger]+
II^\dagger-J^\dagger J=0.$

(ii) For any $\xi=(\xi_1,\xi_2)\in \CC^2\cong \RR^4$ the linear map ${\mathcal D}_\xi\in
\Hom(V\oplus V\oplus W,V\oplus V)$ defined by
\begin{equation}\label{D}
{\mathcal D}_\xi=\left(\begin{array}{lrl}
B_1-\xi_1 & -B_2+\xi_2 & I \\
B_2^\dagger-\bxi_2 & B_1^\dagger-\bxi_1 & J^\dagger 
\end{array}\right)
\end{equation}
is surjective.

The equations $\mu_c=\mu_r=0$ are called the ADHM equations. They are invariant with
respect to the action of the group of unitary transformations of $V$. Solutions of these
equations are called ADHM data. The space of 
ADHM data modulo $U(V)$ transformations has dimension $4rk$ and carries a 
natural hyperk\"ahler metric. ADHM construction identifies this moduli space with the moduli
space of $U(r)$ instantons with $c_2=k$ and fixed trivialization at infinity.
The role of the condition (ii) above is to remove submanifolds in this moduli
space where the hyperk\"ahler metric becomes singular (these are point-like instanton
singularities mentioned in subsection~\ref{instanddbranes}). As a result the moduli space 
of the ADHM data is metrically incomplete.

The instanton connection can be reconstructed from the ADHM data as follows. 
The condition (ii) implies that the family $\Ker {\mathcal D}_\xi$ forms a trivial
subbundle of $V \oplus V\oplus W$ of rank~$r$. Let $v(\xi)$ be its trivialization, i.e.
a linear map $v(\xi): \CC^r\to V \oplus V\oplus W$ smoothly depending on $\xi$ such that
${\mathcal D}_\xi\ v(\xi)=0$ for all $\xi$, and $\rho(\xi)=v(\xi)^\dagger v(\xi)$ is an 
isomorphism for all $\xi$.
We set
$$
A(\xi)=\rho(\xi)^{-1} v(\xi)^\dagger\ dv(\xi). 
$$
The matrix-valued one-form $A$ is a connection on a trivial unitary bundle of rank~$r$.
One can show that its curvature $F_A$ is ASD (see~\cite{Atiyah}). However, it does not
satisfy $A^\dagger=-A$, because we are not using a unitary gauge. Instead
$A$ satisfies
$$
A^\dagger(\xi)=-(\rho(\xi) A(\xi)\rho(\xi)^{-1}+\rho(\xi) d\rho(\xi)^{-1}).
$$
To go to a unitary gauge, we must make a gauge transformation
$$
A'(\xi)=g(\xi) A(\xi) g(\xi)^{-1}+g(\xi) dg(\xi)^{-1},
$$
where $g(\xi)$ is a function taking values in Hermitean $r\times r$ matrices and satisfying
$g(\xi)^2=\rho(\xi).$

We now explain, following~\cite{NS}, how to modify the ADHM construction so that it produces
rank~$r$ instantons on the noncommutative $\RR^4$ defined in the previous section. 
It proves convenient to
apply an orthogonal transformation which brings the matrix $\theta$ in~(\ref{wm}) to
the standard form
$$\theta=\sqrt {-1} \begin{pmatrix}
0 & a & 0 & 0\\
-a & 0 & 0 & 0\\
0 & 0 & 0 & b\\
0 & 0 & -b & 0
\end{pmatrix}.$$
We will assume that $a+b\neq 0.$ Since $\theta$ enters only in the combination
${\h}\theta,$ we can set $a+b=1$ without loss of generality. The relation between
the affine coordinates $\xi_1,\xi_2$ on $\CC^2$ and affine coordinates $x_1,x_2,x_3,x_4$
on $\RR^4$ is chosen as follows:
$$
\xi_1=x_4-\sqrt{-1}\ x_3,\qquad \xi_2=-x_2+\sqrt{-1}\ x_1.
$$
Then $\xi_1,\xi_2,\bxi_1,\bxi_2$ obey the Weyl algebra relations
$$
[\xi_1,\bxi_1]=2{\h}b,\quad [\xi_2,\bxi_2]=2{\h}a, \quad 
[\xi_1,\xi_2]=[\xi_1,\bxi_2]=[\bxi_1,\xi_2]=[\bxi_1,\bxi_2]=0.
$$

The modified ADHM data consist of the same 
four maps which now satisfy $$\mu_c=0,\ \mu_r=-2{\h}(a+b)\cdot 1_{k\times k}.$$ The instanton 
connection is given by essentially the same formulas as in the commutative case.
The operator ${\mathcal D}$ is given by the same formula as ${\mathcal D}_\xi,$ but is now 
regarded as 
an element of $$\Hom_\cA ((V\oplus V\oplus W)\ot_\CC\cA,(V\oplus V)\ot_\CC\cA).$$ 
The module $\Ker {\mathcal D}$ is a projective module over $\cA$. Following~\cite{NB}, 
we assume 
that it is isomorphic to a free module of rank~$r$, and
$v$ is the corresponding isomorphism $v:\cA^{\oplus r}\to \Ker {\mathcal D}.$
We further assume~\cite{NB} that the morphism 
$$
\Delta={\mathcal D}{\mathcal D}^\dagger\in \End_\cA((V\oplus V)\ot \cA)
$$
is an isomorphism.\footnote{One can show that the
latter assumption is always valid provided ${\h}\neq 0.$ As for the former one, it
is not known what constraints should be imposed on the deformed ADHM data to ensure that
$\Ker{\mathcal D}$ is a free $\cA$\!--module of rank $r.$ For $r=1$ 
$\Ker{\mathcal D}$ is never free~\cite{Furu}.} 
Then it is easy to see that
$\rho=v^\dagger v\in \End_{\cA}(\CC^r\ot\cA)$ is an isomorphism too. We set
\begin{equation}\label{asdconnection}
A=\rho^{-1} v^\dagger\ d v.
\end{equation}
(The multiplication here and below is understood to be the Wigner-Moyal multiplication.)
This formula defines a connection 1-form on a trivial unitary bundle on $\cA$ of 
rank~$r$.
The curvature of this connection is given by
$$
F_A=\rho^{-1}dv^\dagger\wedge (1-v\rho^{-1} v^\dagger) dv.
$$
A short computation (essentially the same as in the commutative case) shows that the curvature 
can be written in the form
$$
F_A=\rho^{-1} v^\dagger\ d{\mathcal D}^\dagger\ \Delta^{-1}\ \wedge\ d{\mathcal D}\ v.
$$
Furthermore, since ${\mathcal D}$ and ${\mathcal D}^\dagger$ are linear in $\xi_i,\bxi_i,$ 
their exterior derivatives have a very simple form:
$$
d{\mathcal D}=\begin{pmatrix} -d\xi_1 & d\xi_2 & 0\\
-d\bxi_2 & -d\bxi_1 & 0
\end{pmatrix}, \qquad 
d{\mathcal D}^\dagger=\begin{pmatrix} -d\bxi_1 & -d\xi_2 \\
                                       d\bxi_2 & -d\xi_1 \\
                                       0      & 0 
\end{pmatrix}.
$$ 
Note also that by virtue of the deformed ADHM equations $\Delta$ has a block-diagonal form:
$$
\Delta=\begin{pmatrix}
\delta & 0 \\
0 & \delta
\end{pmatrix},
$$
where $\delta\in \End_\cA(V\ot \cA)$ is an isomorphism. Using this fact, one can easily 
see that $F_A$ is proportional to the 2-forms
$$ d\xi_1\wedge d\bxi_1+d\xi_2\wedge d\bxi_2,\quad d\xi_1\wedge d\bxi_2,
\quad d\xi_2\wedge d\bxi_1,$$
which are anti-self-dual.

As in the commutative case, the connection $A$ does not satisfy $A^\dagger=-A$.
To go to a unitary gauge one has to perform a gauge transformation
$$
A'=g\star A\star g^{-1} + g\star dg^{-1}.
$$
Here $g\in \Aut_\cA(\CC^r\ot \cA)$ should be found from the conditions $g^\dagger=g,$
$g\star g=\rho.$ The existence of such $g$ is an additional assumption.

\subsection{Summary of results}

In the commutative case there is a one-to-one
correspondence between the following three classes of objects:

$A.$ Rank $r$ holomorphic bundles on $\PP^2$ with $c_2=k$ and a fixed trivialization on 
the line at infinity.

$B.$ The set of ADHM data modulo the natural action of $U(k)$.

$C.$ Rank $r$ holomorphic bundles on $\PP^3$ with $c_2=k,$ a trivialization on a fixed line,
vanishing $H^1(E(-2)),$ and satisfying a certain reality condition.

$D.$ $U(r)$ instantons on $\RR^4$ with $c_2=k$.

The correspondence between $C$ and $D$ is a particular instance of twistor 
transform~\cite{AtiyahWard}. The correspondence between $B$ and $C$ has been proved by
Atiyah, Hitchin, Drinfeld, and Manin~\cite{ADHM,Atiyah}. Together these two results
imply that all instantons on $\RR^4$ arise from the ADHM construction. The correspondence 
between $A$ and $B$ has been proved by Donaldson~\cite{Do}. One can also prove the 
correspondence between $A$ and $D$ directly~\cite{Bando,Buchdahl,Guo}. 

The goal of this paper is to extend some of these results to the noncommutative case. We show
that there is a natural one-to-one correspondence between the isomorphism classes of 
the following objects:

$A'.$ Algebraic bundles on a noncommutative deformation of $\PP^2$ with $c_2=k$ and a 
fixed trivialization on the line at infinity.

$B'.$ Deformed ADHM data of Nekrasov and Schwarz modulo the natural $U(k)$ action.

$C'.$ Certain complexes of sheaves on a noncommutative deformation of $\PP^3$ 
satisfying reality conditions.

The moduli space of the deformed ADHM data has a natural hyperk\"ahler metric, and the
other two moduli spaces inherit this metric.

Furthermore, we reinterpret the deformed ADHM construction of Nekrasov and
Schwarz in terms of a noncommutative deformation of the twistor transform.

It is interesting to note that H.~Nakajima~\cite{Na2} studied the same linear algebra data as 
Nekrasov and Schwarz and showed that their moduli space coincides with the moduli space of 
torsion free sheaves on a commutative $\PP^2$ with a trivialization on a fixed line. On the 
other hand, we show that the same data describe algebraic bundles on a noncommutative $\PP^2$. 
As shown below, the interpretation in terms of complexes of sheaves on a noncommutative $\PP^3$ 
provides a geometric reason for this ``coincidence.'' We prove that the two moduli spaces
are isomorphic as hyperk\"ahler manifolds, but the natural complex structures on them
differ by an $\SO(3)$ rotation.

The rest of the paper is organized as follows.  In Section~\ref{varieties}
we define noncommutative deformations of certain commutative projective varieties
($\PP^2,$ $\PP^3,$ and a quadric in $\P^5$). Section~\ref{algebra} is an algebraic
preparation for the study of bundles on noncommutative projective spaces.
In Section~\ref{sheaves} we study the cohomological properties of sheaves on
noncommutative $\PP^2$ and $\PP^3$ and define locally free sheaves (i.e. bundles).
In Section~\ref{bundles} we show that any bundle on a noncommutative $\PP^2$ trivial
on the commutative line at infinity arises as a cohomology of a monad. 
In Section~\ref{pthree} we exhibit bijections between 
$A',$ $B',$ and $C'$ and explain the relation with Nakajima's results. 
In Section~\ref{twistor} we construct a noncommutative deformation of 
Grassmannians and flag manifolds and describe a noncommutative version of the twistor transform. 
We also describe a nice class of noncommutative projective varieties associated with a 
Yang-Baxter operator and define differential forms on these varieties. 
In section~\ref{sec:qdeformed} we consider a more general deformation
of $\RR^4$ (a $q$\!--\,deformed $\RR^4$) whose physical significance is obscure at present. 
We propose an ADHM--like construction of instantons
on this space and outline its relation to noncommutative algebraic geometry.  
In the Appendix we define
the Wigner-Moyal product on the space of $C^\infty$ functions on $\RR^n$ all of whose 
derivatives are polynomially bounded, and prove that the Wigner-Moyal product provides 
this space with a structure of an algebra over $\CC.$

\section{Geometry of noncommutative varieties}\label{varieties}

\subsection{Algebraic preliminaries}

Let $\kk$ be a base field (we will be dealing only with $\kk=\C$ or $\kk=\R$ in this paper).
Let $A$ be an algebra over $\kk$. It is called right (left) noetherian if 
every right (left) ideal is finitely generated, and it is called noetherian 
if it is both right and left noetherian.

Let $A=\mathop{\op}\limits_{i\ge 0} A_{i}$ be a graded noetherian algebra.
We denote by $mod(A)$ the category of finitely generated right $A$\!--modules,
by $gr(A)$ the category of finitely generated graded right $A$\!--modules, and 
by $tors(A)$ the full subcategory of $gr(A)$ which consists of finite 
dimensional graded $A$\!--modules.
 
An important role will be played by the quotient category 
$qgr(A)=gr(A)/tors(A)$. It has the following explicit description.
The objects of $qgr(A)$ are the objects of $gr(A)$ (we denote by $\widetilde{M}$
the object in $qgr(A)$ which corresponds to a module $M$). The morphisms
in $qgr(A)$ are given by 
$$
\Hom_{qgr}(\widetilde{M}, \widetilde{N})=\lim_{\stackrel{\lto}{M'}} \Hom_{gr}(M', N)
$$
where $M'$ runs over submodules of $M$ such that $M/M'$ is finite dimensional.

On the category $gr(A)$ there is a shift functor: for a given graded module
$M=\mathop{\op}_{i\ge 0} M_{i}$ the shifted module $M(r)$ is defined by 
$M(r)_{i}= M_{r+i}$.
The induced  shift functor on the quotient category $qgr(A)$ 
sends $\widetilde{M}$ to $\widetilde{M}(r)=\widetilde{M(r)}$.

Similarly, we can consider the category $Gr(A)$ of all graded right
$A$\!--modules. It contains the subcategory $Tors(A)$ of torsion modules.
Recall that a module $M$ is called torsion if for any element $x\in M$ one has
$x A_{\ge s}=0$ for some $s,$ where $A_{\ge s}=
\mathop{\op}\limits_{i\ge s} A_{i}$.
We denote by $QGr(A)$ the quotient category $Gr(A)/Tors(A)$.
The category $QGr(A)$ contains $qgr(A)$ as a full subcategory.
Sometimes it is convenient to work in $QGr(A)$ instead of $qgr(A)$.

Henceforth, all graded algebras will be noetherian algebras generated by
the first component $A_{1}$ with $A_{0}=\kk$.

Sometimes we use subscripts $R$ or $L$ for categories $gr(A),\ qgr(A),$ etc.,
to specify whether right or left modules are considered.
If the subscript is omitted, the modules are taken to be right modules.
For the same reason for an $A$\!--bimodule $M$ we sometimes write $M_{A}$
or ${}_{A} M$ to specify whether the right or left module structure is considered.

\subsection{Noncommutative varieties}

A variety in commutative geometry is a topological space with a sheaf
of functions (continuous, smooth, analytic, algebraic, etc.) which is, obviously, 
a sheaf of algebras.
One of the main objects in geometry (algebraic or differential) is a bundle
or, more generally, a sheaf. 
To any variety $X$ we can associate an abelian category
of sheaves of modules (maybe with some additional properties) 
over the sheaf of algebras of functions. Given a sheaf of modules on $X,$ the 
space of its global sections is a module over the algebra of global functions on $X.$ 
Thus the functor of global sections associates to every $X$ an algebra and a certain category
of modules over it.
Under favorable circumstances, much of the information about the geometry of $X$ is 
contained in this purely algebraic datum. Let us give a few examples.

If $X$ is a compact Hausdorff topological space, then the category of vector bundles over $X$
is equivalent to the category of finitely generated projective modules over the algebra of 
continuous functions on $X$~\cite{Serre,Swan}. The equivalence is given by the functor
which maps a vector bundle to the module of its global sections.

It is well known that if $A$ is a commutative noetherian algebra, the category of 
coherent sheaves on the noetherian affine scheme $Spec(A)$ is equivalent to the category of 
finitely generated modules over $A$. The equivalence is again given by the functor which 
attaches to a coherent sheaf the module of its global sections.

In the case of projective varieties the only global functions are constants, 
so one has to act somewhat differently. Since a projective variety $X$ is by definition
a subvariety of a projective space, it inherits from it the line bundle $\O_X(1)$
and its tensor powers $\O_X(i)$.
We can consider a graded algebra 
$$
\Gamma(X)=\mathop{\oplus}\limits_{i\ge 0} H^{0}(X,\O_X(i)).
$$

This algebra is called the homogeneous coordinate algebra of $X$.
Furthermore, for any sheaf $\F$ we can define a graded $A$\!--module 
$$
\Gamma(\F)= \mathop{\oplus}\limits_{i\ge 0} H^{0}(X,\F(i)).
$$
It can be checked that $\Gamma$ is a functor from the category of coherent sheaves on $X$ 
$coh(X)$ to $gr(\Gamma(X))$.
In a brilliant paper \cite{Se}, J-P.~Serre described the category of coherent sheaves 
on a projective scheme $X$ in terms
of graded modules over the graded algebra $\Gamma(X)$. 
He proved that the category $coh(X)$ is equivalent to the quotient category
$qgr(\Gamma(X))=gr(\Gamma(X))/tors(\Gamma(X))$.
The equivalence is given by the composition of the functor $\Gamma$ with the projection
$\pi: gr(A)\to qgr(A)$.
On other hand, let $A=\mathop{\op}\limits_{i\ge 0} A_{i}$ be a graded commutative algebra 
generated over $\kk$ by 
the first component (which is assumed to be finite dimensional). We can associate to $A$
a projective scheme $X=Proj(A)$. Serre proved that the category $coh(X)$ is equivalent
to the category $qgr(A)$.

The equivalence also holds for the category 
of quasicoherent sheaves on $X$ and the category $QGr(A)=Gr(A)/Tors(A)$.

In all of the above examples it turned out that the natural category of sheaves or 
bundles on a variety is equivalent to a certain category defined in terms of (graded) modules
over some (graded) algebra.
On the other hand, ``as A.~Grothendieck taught us, to do geometry you really don't need a 
space, all you need is a category of sheaves on this would-be space'' 
(\cite{Ma2}, p.83).

For this reason, in the realm of algebraic geometry it is natural to
regard a noncommutative noetherian algebra as a coordinate algebra of a noncommutative 
affine variety; 
then the category of finitely generated right modules over this algebra is 
identified with the category of coherent sheaves on the corresponding variety.
Similarly, a noncommutative graded noetherian algebra is regarded as a homogeneous
coordinate algebra of a
noncommutative projective variety. The category of finitely generated graded 
right modules over this algebra 
modulo torsion modules is identified with
the category of coherent sheaves on this variety 
(see \cite{Ar}, \cite{Ma2}, \cite{SV}).

A different approach to noncommutative geometry has been pursued by A.~Connes~\cite{Connes}.

\subsection{Noncommutative deformations of commutative varieties}

Many important noncommutative varieties arise as deformations of 
commutative ones.

Let $X$ be a commutative variety (affine or projective).
Let $A$ be the corresponding commutative (graded) algebra.
A noncommutative deformation
of $X$ is a deformation of the algebra structure on $A,$
that is, a deformation of the multiplication law. 
Usually it is not easy to write down an explicit formula for the
deformed product. 

There is a more algebraic way to describe noncommutative deformations of commutative
varieties. Assume that the algebra $A$ is given in terms of generators and relations. 
This means that $A$
is given as a quotient $A=T(V)/\langle R\rangle,$ where $V$ is the vector space
spanned by the generators, $T(V)$ is the
tensor algebra of $V,$ and $\langle R\rangle$ is a 
two-sided ideal in $T(V)$ generated by a subspace of relations $R\subset T(V)$.
Assume that $R_\h\subset T(V)$ is a one-parameter deformation of the
subspace $R$. Then $A_\h=T(V)/\langle R_\h\rangle$ is 
a one-parameter deformation of $A$.
(If $A$ is graded, then we assume that $R$ is a graded subspace,
and the deformation preserves the grading).

We denote by $X_\h$ the noncommutative variety corresponding
to the algebra $A_\h$. Thus $X_\h$ is a noncommutative 
one-parameter deformation of $X$.

If $X$ is projective and $A$ is a graded algebra, then we denote by 
$coh(X_{\h})$ the category $qgr(A_{\h})$. Furthermore, as in the commutative 
case, we will write $\O(r)$ for the object $\widetilde{A_{\h}}(r)$.

Now we define noncommutative varieties which are going to be used in this
paper.

\subsection{Noncommutative $\C^4$}\label{noncomC}

Denote by $A(\C^4)$  the algebra of polynomial
functions on $\C^4$. Let $\theta$ be a skew-symmetric
$4\times 4$ matrix.

Let us define the algebra $A(\C_{\h}^4)$ as
an algebra over $\C$ generated by $x_i$ ($i=1,2,3,4$) with relations
$[x_i,x_j]=\h\theta_{ij}$:
\begin{equation}\label{cfour}
A(\C_{\h}^4) = \T(x_1,x_2,x_3,x_4)/\langle[x_i,x_j]=
\h\theta_{ij}\rangle _{1\le i,j \le 4}.
\end{equation}

We will regard $A(\C_{\h}^4)$ as the algebra of polynomial functions
on a noncommutative affine variety $\C_{\h}^4$. 

\subsection{Noncommutative 4-dimensional quadric}

Let $G$ be a $4\times 4$ symmetric nondegenerate matrix.
Consider a graded algebra  $Q_\h = \mathop{\op}\limits_{i\ge 0} Q_i$ over $\C$
generated by the elements $X_1,X_2,X_3,X_4,D,T$ of degree $1$ with
the following quadratic relations:
\begin{equation}\label{quadrel}
\begin{array}{ll}
{}[T, D]=[T,X_i]=0,\\
{}[X_{i}, X_{j}]=\h \theta_{ij} T^2,\\
{}[D, X_{i}]= 2\h \sum\limits_{lk} \theta_{il} G^{lk} X_k T,\\
\sum\limits_{ij} G^{ij}X_{i}X_{j}=DT.
\end{array}
\end{equation}

We denote by $\Q^4_\h$ the noncommutative projective variety 
corresponding to the algebra $Q_\h$. It is evident that
$\Q^4_\h$ is a deformation of a 4-dimensional commutative 
quadric $\Q^4=\{\sum_{ij}G^{ij}X_iX_j=DT\} \subset \C\P^5$.

\subsection{Embedding $\C^4_\h\hookrightarrow\Q^4_\h$}

Let $Q_{\h}[T^{-1}]$ be the localization of the algebra $Q_{\h}$ with respect to $T$.
Elements of degree $0$ in $Q_{\h}[T^{-1}]$ form a subalgebra which will be denoted by 
$Q_{\h}[T^{-1}]_{0}$. 
\begin{lemma}
The map $x_i\mapsto T^{-1}X_i$ $(i=1,2,3,4)$ induces an
isomorphism of the algebra $A(\C_{\h}^4)$ with the algebra $Q_{\h}[T^{-1}]_{0}$.
\end{lemma}
\begin{proof} Obvious.
\end{proof}

This means that  $\C^4_\h$ can be identified with the open subset $\{T\ne0\}$
in $\Q^4_\h$. For this reason, $\Q^4_\h$ may be regarded as a 
compactification of $\C^4_{\h}$ which is compatible with the bilinear form~$G$.
Note also that the complement of $\C^4_\h$
in $\Q^4_\h$ corresponds to the algebra
$$
Q_\h/\langle T\rangle  = 
\T(X_1,X_2,X_3,X_4,D)/\langle [X_i,X_j]=[D,X_i]=0,\ \sum_{ij}G^{ij}X_iX_j=0
\rangle .
$$
Since this algebra is commutative, the complement is the usual 
3-dimensional commutative quadratic cone. Thus one may say that $\Q^4_{\h}$ is obtained from
$\C^4_\h$ by adding a cone ``at infinity''. This is in complete analogy
with the commutative case.

\subsection{Noncommutative $\Plh$ and $\PTh$}

Noncommutative deformations of the projective plane have been classified in 
\cite{AS}, \cite{ATV},
\cite{BP}. We will need one of them,
namely the one whose homogeneous coordinate algebra is a graded algebra 
$\plh=\mathop{\op}\limits_{i\ge 0}\plh_{i}$ 
over $\C$  generated by the elements $w_{1}, w_{2}, w_{3}$ of degree 1 with 
the relations:
\begin{equation}\label{re}
\begin{array}{l}
{}[w_3, w_i]=0\ \mbox{ for any } \ i=1,2,3,\\
{}[w_1, w_2]= 2\h w_{3}^{2}.
\end{array}
\end{equation}

We will also need a noncommutative deformation of the 3--dimensional projective space, whose
homogeneous coordinate algebra will be denoted $\pth=\mathop{\op}\limits_{i\ge 0} \pth_{i}$. 
It is a graded algebra over $\CC$ generated by $\pth_1=U,$ where the vector space $U$ is
spanned by elements $z_1,z_2,z_3,z_4$ obeying the relations
\begin{equation}\label{rel}
\begin{array}{l}
{}[z_3, z_i]=[z_4,z_i]=0\ \mbox{ for any } \ i=1,2,3,4,\\
{}[z_1, z_2]= 2\h z_{3} z_{4}.
\end{array}
\end{equation}

The noncommutative projective varieties corresponding to $\plh$ and $\pth$ will be denoted 
$\Plh$ and $\PTh,$ respectively.

Note that for $\h\ne 0$ all algebras $\pth$ are isomorphic, and therefore the varieties
$\PTh$ are the same for all $\h\ne 0$. The same is true for $\Plh.$

\subsection{Subvarieties in $\PTh$ and $\Plh$}\label{subvarieties}

If $I\subset\pth$ is a graded two-sided ideal, then the quotient algebra $\pth/I$
corresponds to a closed subvariety $X(I)\subset \PTh$. Let us describe some of them.

Let $J$ be the graded two-sided ideal generated by $z_3$ and $z_4$. 
Then $\pth/J=\T(z_1,z_2)/\langle [z_1,z_2]=0\rangle ,$ hence $X(J)$ is the commutative 
projective line.

For each point $p=(\lambda:\mu)\in\P^1$ let $J_p$ denote the graded two-sided 
ideal generated by $\lambda z_3+\mu z_4$. If $p=(0:1)$ or $p=(1:0),$ then
it is easy to see that $X(J_p)$ is the commutative projective plane. 
For all other $p\in\P^1$ we have
$$
\pth/J_p=\T(z_1,z_2,z_3)/
\left\langle [z_1,z_3]=[z_2,z_3]=0,\ [z_1,z_2]=-2\h\frac\lambda\mu z_3^2\right\rangle ,
$$
hence $X(J_p)$ is a noncommutative projective plane isomorphic to $\Plh$.

We have $J_p\subset J$ for all $p\in\P^1,$ hence all planes $X(J_p)$ pass through
the line $X(J)$. Thus we see that  {\em $\PTh$ is a pencil of noncommutative 
projective planes passing through a fixed commutative projective line}.

Similarly, the two-sided ideal generated by $w_3$ in $\plh$ corresponds to a 
commutative projective line 
$l=\{w_3=0\}\subset \Plh$.

\section{Properties of algebras $\pth$ and $\plh$ and
the resolution of the diagonal}\label{algebra}

This section is a preparation for the study of sheaves on $\PTh$ and $\Plh$.
We show that the algebras $\pth$ and $\plh$ are  regular and Koszul and construct
the resolution of the diagonal, which will enable us to associate  monads
to certain bundles on $\Plh$.

\subsection{Quadratic algebras}

A graded algebra $A=\mathop{\op}\limits_{i\ge 0}A_{i}$ over a field $\kk$
is called quadratic if it is connected (i.e. $A_{0}=\kk$),
is generated by the first component $A_{1},$ and the ideal of
relations is generated by the subspace of quadratic
relations $R(A)\subset A_{1}\ot A_{1}$.

Therefore the algebra $A$ can be represented as $T(A_{1})/\langle R(A)\rangle,$
where $T(A_{1})$ is a free tensor algebra generated by the space $A_{1}$.

The algebras $\pth$ and $\plh$ are  quadratic algebras.
For example, $\pth$ can be represented 
as $\T(\Vec)/\langle W\rangle,$ 
where $\Vec=\pth_{1}$ is a 4-dimensional vector space and $W$ is the 
6--dimensional subspace
of $\Vec\ot \Vec$ spanned by the relations (\ref{rel}).

\subsection{The dual algebra}

For any quadratic algebra $A= T(A_{1})/\langle R(A)\rangle$ we can define 
its dual algebra which is also quadratic.

Let us identify $A_{1}^*\ot A_{1}^*$ with 
$(A_{1}\ot A_{1})^*$ by $(l\ot m)(a\ot b)= m(a)l(b)$.
Denote by  $R(A)^{\perp}$ the annulator of $R(A)$ in 
$A_{1}^*\ot A_{1}^*,$ i.e. the subspace which consists 
of such $q\in (A_{1}^{*})^{\ot 2}$ that 
$q(r)=0$ for any
$r\in R(A)$.  
\begin{definition}{\rm (\cite{Ma2})}
The algebra $A^{!}=T(A_{1}^*)/\langle R(A)^{\perp}\rangle$
is called the dual algebra of $A$.
\end{definition}
\begin{example}
Let $\{\cz_{i}\}, i=1,2,3,4,$ be the basis of $\pth^{!}_{1}=\Vec^*$ 
which is dual to
$\{z_{i}\}$. By definition, $\pth^!$  is generated by 
$\{\cz_{i}\}$ with
defining relations
$$
\begin{array}{l}
\cz_{i}^2=0 \quad \mbox{ for all} \ i=1,...,4;\\
 \cz_{i}\cz_{j}+\cz_{j}\cz_{i}=0 
 \quad \mbox{ for all} \ i<j, (i,j)\ne (3,4);\\
\cz_3 \cz_{4}+ \cz_{4}\cz_{3}=\h[\cz_{1}, \cz_{2}]=
2\h\cz_{1}\cz_{2}.
\end{array}
$$
\end{example}

In the commutative case the dual algebra of the symmetric algebra $S^{\cdot}(\Vec)$ 
is isomorphic
to the exterior algebra $\Lambda^{\cdot}(\Vec^*)$.
Obviously, the algebras $\pth^!$ and $\plh^!$ are deformations of 
exterior algebras. 
For example, the vector space $\pth^!_{k}$ is spanned
by the elements $ \cz_{i_1}\cdots\cz_{i_k}$ with $i_{1}<\cdots<i_{k}$.
In particular, the dimension of the vector space $\pth^!_{k}$ is equal to
$\binom{4}{k}$. Similarly, the dimension of $\plh^!_{k}$ is equal to
$\binom{3}{k}$.
\begin{proposition} Let $A$ be $\pth$ or $\plh,$ and let $n$ be
$4$ or $3,$ respectively.
The multiplication map
$A^!_{k}\ot A^!_{n-k}\lto A^!_{n}$ is a non-degenerate pairing.
Hence the dual algebra $A^!$ is a Frobenius algebra, i.e. 
$(A^!)_{A^!}\cong ({}_{A^!}A^!)^*$
as right $A^!$--modules.
\end{proposition}
\begin{proof} 
The proposition holds for the exterior algebra,
and therefore also for the algebra $A^!,$ since the latter is a ``small'' deformation of 
the exterior algebra.\hfill
\end{proof}

\subsection{The Koszul complex}

Consider right $A$--modules $(A^{!}_{k})^* \ot A$. 
The following complex $K_{\cdot}(A)$ is called 
the (right) Koszul complex of a quadratic algebra:
$$
\cdots
\stackrel{d}{\lto}(A^{!}_{3})^* \ot A(-3)
\stackrel{d}{\lto}(A^{!}_{2})^* \ot A(-2)
\stackrel{d}{\lto}(A^{!}_{1})^* \ot A(-1)
\stackrel{d}{\lto}(A^{!}_{0})^* \ot A\lto 0, 
$$
where the map $d:(A^{!}_{k})^* \ot A\to (A^{!}_{k-1})^* \ot A$ is  a composition 
of two natural maps:
$$
(A^{!}_{k})^* \ot A\lto (A^{!}_{k})^*\ot A_{1}^{!}\ot A_{1}\ot A
\lto(A^{!}_{k})^* \ot A.
$$
Here the first arrow sends $\alpha\ot a$ to $\alpha\ot e\ot a$
with $e$ defined as
$$
e=\sum_{i} y_{i}\ot x_{i} \in A^!_{1}\ot A_{1},
$$ 
and $\{x_{i}\}$ and $\{y_{i}\}$ being the dual bases of $A_{1}$ and $A_{1}^!,$
respectively.
The second map is determined by the algebra structures on $A^!$ and $A$.

It is a well--known fact that $d^2=0$ (see, for example, \cite{Ma2}).

Let $\kk_{A}$ be the trivial right $A$-module. The Koszul complex $K_{\cdot}(A)$
possesses a natural augmentation 
$K_{\cdot}\stackrel{\varepsilon}{\lto}\kk_{A}\lto 0$.
\begin{definition}{\rm (see \cite{Pr})} A quadratic algebra $A=\mathop{\op}\limits_{i\ge 0} A_{i}$
is called a Koszul algebra if the augmented Koszul complex
$K_{\cdot}(A)\stackrel{\varepsilon}{\lto}\kk_{A}\lto 0$ is exact. 
\end{definition}
In the same manner one can define the left Koszul complex of a quadratic algebra.
It is well known that the exactness of the right Koszul complex is equivalent to the exactness
of the left Koszul complex (see, for example, \cite{Lo}).
\begin{proposition}
The algebras $\pth$ and $\plh$  are Koszul algebras.
\end{proposition}
\begin{proof} For $\h=0$ this is a well-known fact about the symmetric algebra 
$S^{\cdot}(\Vec)$. Since the augmented Koszul complex is exact for $\h=0,$
it is also exact for small $\h,$ and consequently for all $\h$.
\hfill\end{proof}

Since the dual algebras $\pth^!$ and $\plh^!$ are finite,
the Koszul resolutions for the algebras $\pth$ and $\plh$ are finite too and 
have the same form as the resolutions for ordinary symmetric algebras.
For example, the Koszul resolution for $A=\plh$ is:
$$
\{ 0\to
(A^{!}_{3})^* \ot A(-3)
\to(A^{!}_{2})^* \ot A(-2)
\to(A^{!}_{1})^* \ot A(-1)
\to (A^{!}_{0})^* \ot A \}\to \C.
$$

\subsection{Resolution of the diagonal}

Consider a bigraded vector space 
$$K_{\twice}^2(A)=\bigoplus\limits_{k,l\ge0}K^2_{k,l}(A)
\quad
\text{with}
\quad
K^2_{k,l}(A)=A(k)\ot (A^!_{l-k})^*\ot A(-l).
$$
Consider morphisms $d_R:K^2_{k,l}\to K^2_{k,l-1}$ and $d_L:K^2_{k,l}\to 
K^2_{k+1,l}$
given by the following compositions 
$$
\begin{array}{l}
d_R: A\ot (A^!_k)^*\ot A \to A\ot (A^!_k)^*\ot A^!_1\ot A_1 \ot A \to A\ot 
(A^!_{k-1})^*\ot A,\\
d_L: A\ot (A^!_k)^*\ot A \to A\ot A_1\ot A^!_1 \ot (A^!_k)^*\ot A \to A\ot 
(A^!_{k-1})^*\ot A.
\end{array}
$$
Here the leftmost maps are given by 
$$
e_R = \sum_i y_i\ot x_i \in A^!_1\ot A_1
\quad\mbox{and}\quad
e_L = \sum_i x_i\ot y_i \in A_1\ot A^!_1,
$$
where $\{x_i\}$ and $\{y_i\}$ are the dual bases of $A_1$ and $A^!_1,$
respectively,
while the rightmost maps are induced by the algebra structures of $A^!$ and $A$.
It is easy to show that 
$$
d_R^2=d_L^2=0\qquad\mbox{and}\qquad d_Rd_L=d_Ld_R,
$$
hence $K_{\twice}^2(A)$ is a bicomplex. It is called the {\em double Koszul bicomplex} 
of the quadratic algebra~$A$.

The topmost part of the bicomplex looks as follows:
$$
\begin{CD}
\dots @>{d_R}>> A\ot (A^!_{l+1})^*\ot A(-1-l) @>{d_R}>> A\ot (A^!_l)^*\ot A(-l)  
   @>{d_R}>> \dots \\
@.	        @V{d_L}VV			  @V{d_L}VV				
  \\
\dots @>{d_R}>> A(1)\ot (A^!_l)^*\ot A(-1-l)     @>{d_R}>> A(1)\ot 
(A^!_{l-1})^*\ot A(-l) @>{d_R}>> \dots
\end{CD}
$$

Each term of the bicomplex $K_{\twice}^2(A)$ has an obvious structure
of a bigraded $A$-bimodule, and it is clear that the 
differentials are morphisms of bigraded $A$-bimodules. 

Let
$$
\CK_l(A) = \operatorname{Ker} d_L:K^2_{0,l}(A) \to K^2_{1,l}(A).
$$
Then $\CK_{\cdot}(A)$ is a complex of bigraded 
$A$-bimodules (with respect to the differential 
$d_R$).

Consider a bigraded algebra $\Delta=\bigoplus_{i,j}\Delta_{ij}$ 
with $\Delta_{ij}=A_{i+j}$ and with the multiplication induced
from $A$. The algebra $\Delta$ is called the
diagonal bigraded algebra of $A$.
Note that the multiplication map induces a surjective
morphism of $A$-bimodules $\delta:A\ot A\to \Delta$.

\begin{lemma}
The map
$$
\delta:\CK_0(A) = A\ot A \to \Delta
$$
gives an augmentation of the complex $\CK_{\cdot}(A)$.
\end{lemma}
\begin{proof}
We have to check that $\delta \cdot d_R:\CK_1(A)\to A$ vanishes.
Note that $K^2_{0,1}(A)=A\ot A_1\ot A(-1),$ and that the differentials
$d_R$ and $d_L$ restricted to $K^2_{0,1}(A)$ coincide with 
the multiplication maps $m_{1,2}$ and $m_{2,3},$ respectively.
Thus we have the following commutative diagram:
$$
\begin{CD}
\CK_1(A)      @>{d_R}>>     \CK_0(A) @>{\delta}>> \Delta  \\
@VVV                        @|             @| \\
A\ot A_1\ot A(-1) @>{m_{1,2}}>> A\ot A   @>{\delta}>> \Delta  \\
@V{m_{2,3}}VV \\
A(1)\ot A(-1)
\end{CD}
$$
Now the Lemma follows because $\delta\cdot m_{1,2} = \delta\cdot m_{2,3}$ 
(associativity)
obviously annihilates $\Ker m_{2,3} = \CK_1(A)$.
\hfill\end{proof}

\begin{proposition}\label{ckexact}
If $A$ is Koszul, then $\CK_\cdot(A) \stackrel{\delta}{\to} \Delta$ is exact.
\end{proposition}
\begin{proof}
The $(p,q)$\!--bigraded component of $K^2_{k,l}(A)$ is equal to 
$A_{p+k}\ot(A^!_{l-k})^*\ot A_{q-l},$ hence the $(p,q)$\!--
bigraded component of the bicomplex $K_{\twice}^2(A)$ vanishes for $l<k$ or $l>q$.
Thus the $(p,q)$\!--bigraded component of the bicomplex $K_{\twice}^2(A)$ is bounded.
Therefore both spectral sequences of the bicomplex $K_{\twice}^2(A)$ converge
to the cohomology of the total complex $\operatorname{Tot}(K_\twice^2(A))$.
The first term of the first spectral sequence reads
$$
E^1_{k,l}=\begin{cases} A(l)\ot \kk(-l), & \text{if $k=l$}\\ 0, & 
\text{otherwise}\end{cases}
$$
Hence the spectral sequence degenerates in the first term, and we have
$$
H^0(\operatorname{Tot}(K_{\twice}^2(A))) = \bigoplus_{l=0}^\infty A(l)\ot 
\kk(-l),\qquad
H^{\ne0}(\operatorname{Tot}(K_\twice^2(A))) = 0.
$$
On the other hand, the first term of the second spectral sequence reads
$$
E^1_{k,l}=\begin{cases} 
\kk(l)\ot A(-l), & \text{if $k=l>0$}\\ 
\CK_l(A), & \text{if $k=0$}\\
0, & \text{otherwise}
\end{cases}
$$
Hence the spectral sequence degenerates in the second term, and we have
$$
H^0(\operatorname{Tot}(K_\twice^2(A))) = H^0(\CK_\cdot(A))\oplus\left(\bigoplus_{l=1}^\infty 
\kk(l)\ot A(-l)\right),
\qquad H^{l}(\operatorname{Tot}(K_\twice^2(A))) = H^l(\CK_\cdot(A)).
$$
Therefore $H^{\ne0}(\CK_\cdot(A))=0,$ and we have an exact sequence
$$
0 \to H^0(\CK_\cdot(A)) \to \bigoplus_{l=0}^\infty A(l)\ot \kk(-l) \to 
\bigoplus_{l=1}^\infty \kk(l)\ot A(-l) \to 0.
$$
Looking at $(p,q)$\!--bigraded component of this sequence we see that 
$$
(H^0(\CK_\cdot(A)))_{p,q} = \begin{cases} A_{p+q}, & \text{if $p,q\ge 0$}\\ 0, & 
\text{otherwise}\end{cases}
$$
Thus $H^0(\CK_\cdot(A))=\Delta$.
\hfill\end{proof}

\begin{definition}\label{Omk}
Define the left $A$\!--module  
$\Om^k$ as the cohomology of the left Koszul complex, 
truncated in the term $K_k$. In particular, $\Om^1$ is defined by the
so-called  Euler sequence
\begin{equation}\label{euler}
0 \to \Om^1 \to A(-1)\ot A_1 \stackrel m\to A \stackrel\varepsilon\to \kk \to 0.
\end{equation}
\end{definition}
In section \ref{differentials} we will show that for noncommutative projective spaces
the sheaves corresponding to the modules
$\Om^k$ can be regarded as sheaves of differential forms.

\begin{proposition}\label{ckequals}
We have $\CK_k(A) = \Om^k(k)\ot A(-k)$.
\end{proposition}
\begin{proof}
This follows immediately from the definition of $\Om^k$ and $\CK_{k}(A)$.
\hfill\end{proof}

Combining Propositions \ref{ckexact} and \ref{ckequals}, 
we obtain the following resolution of the diagonal:
\begin{equation}\label{diagonal}
\dots \lto \Om^2(2)\ot A(-2) \lto \Om^1(1)\ot A(-1) \lto A\ot A 
\lto \Delta \lto 
0.
\end{equation}

\subsection{Cohomological properties of the algebras $\pth$ and $\plh$}

First we note that both algebras $\pth$ and $\plh$ are noetherian.
This follows from the fact that they are Ore extensions of commutative
polynomial algebras (see for example, \cite{MR}).
For the same reason the algebras $\pth$ and $\plh$ have finite right (and left)
global dimension, which is equal to 4 and 3, respectively (see \cite{MR}, p.~273).

We remind that the global dimension of a ring $A$
is the minimal number $n$ (if it exists)
such that for any two modules $M$ an $N$ we have
$\Ext^{n+1}_{A}(M, N)=0$.

In the paper \cite{AS} the notion of a regular algebra has been introduced.
Regular algebras have many good properties (see \cite{Ar}, \cite{ATV}, \cite{YZ}, etc.).
\begin{definition}\label{reg}
 A graded algebra $A$ is called regular of dimension $d$
if it satisfies the following conditions:

\begin{tabular}{ll}
{\rm(1)}& $A$ has global dimension $d,$\\
{\rm(2)}& $A$ has polynomial growth, i.e. $\dim A_n\le cn^{\delta}$ for some 
$c, \delta\in \R,$\\
{\rm(3)}& $A$ is Gorenstein, meaning that $\Ext^{i}_{A}(\kk, A)=0$ if $i\ne d,$\\
& and 
$\Ext^{d}_{A}(\kk, A)=\kk(l)$ for some $l$.
\end{tabular}
\end{definition}
Here $\Ext_{A}$ stands for the Ext functor in the category $mod(A)$.

It is easy to see that these properties are verified for $\pth$ and $\plh$.
Property (2) holds because our algebras grow as ordinary
polynomial algebras. Property (3) follows from the fact that $\pth$ and $\plh$
are Koszul algebras and the dual algebras are Frobenius resolutions.
In this case the Gorenstein parameter $l$ in (3) is equal to the global dimension
$d$.
Thus we have
\begin{proposition} The algebras $\pth$ and $\plh$ are noetherian regular
algebras of global dimension $4$ and $3,$ respectively.
For these algebras the Gorenstein parameter $l$ coincides with the global dimension $d$.
\end{proposition}

\section{Cohomological properties of sheaves on $\Plh$ and $\PTh$}\label{sheaves}

\subsection{Ampleness and cohomology of $\O(i)$}

Let $A$ be a graded algebra and $X$ be the corresponding noncommutative projective variety.
Consider the sequence of sheaves $\{ \O(i) \}_{i\in \Z}$ in the category 
$coh(X)\cong qgr(A),$ where $\O(i)= \widetilde{A(i)}$. 

This sequence 
is called ample if the following conditions hold:

\begin{tabular}{ll}
(a)& For every coherent sheaf $\F$ there are integers $k_{1},..., k_{s}$ and
an epimorphism \\
&$\mathop{\op}\limits_{i=1}^{s} \O(-k_{i})\lto \F$.\\

(b)& For every epimorphism $\F\lto \G$ the induced map\\
&$\Hom(\O(-n), \F)\lto \Hom(\O(-n), \G)$ is surjective for $n\gg 0$.\\
\end{tabular}
\bigskip

It is proved in \cite{Ar} that the sequence $\{ \O(i) \}$ is ample in $qgr(A)$
for a graded right noetherian $\kk$--algebra $A$
if it satisfies the extra condition:
$$
(\chi_{1})  : \quad \dim_{\kk}\Ext^{1}_{A}(\kk, M) < \infty 
$$
for any finitely generated graded $A$\!--module $M$.

This condition can be verified for all noetherian
regular algebras (see \cite{Ar}, Theorem 8.1). In particular, the categories
$coh(\PTh),$ $coh(\Plh)$  have ample sequences.

For any sheaf $\F\in qgr(A)$ we can define a graded module $\Gamma(\F)$ by the rule:
$$
\Gamma(\F):= \mathop{\oplus}\limits_{i\ge 0} \Hom(\O(i), \F)
$$
It is proved in \cite{Ar} that 
for any noetherian algebra $A$ that satisfies the condition $\chi_1$
the correspondence $\Gamma$ is a functor from $qgr(A)$ to $gr(A)$
and the composition of $\Gamma$ with the natural projection
$\pi : gr(A)\lto gqr(A)$ is isomorphic to the identity functor
(see \cite{Ar}, ch.~3,4). 

Now we formulate a result about the cohomology of sheaves on
noncommutative projective spaces. This result is proved in \cite{Ar}
for a general regular algebra and parallels the commutative case.

\begin{proposition}{\rm (Theorem 8.1. \cite{Ar})}
 Let $A$ be $\pth$ or $\plh,$ and $X$ be $\PTh$ or
$\Plh,$ respectively.
Denote by $n$ the dimension of $X$ (in our case $n=3$ or $n=2,$ respectively). 
Then 

{\rm 1)} The cohomological dimension of $coh(X)$ is equal to $\dim(X),$
i.e. for any two coherent sheaves $\F$ and $\G$ $\Ext^{i}(\F, \G)$ vanishes if
$i>n$.

{\rm 2)} There are isomorphisms
\begin{equation}\label{cohinv}
H^{p}(X, \O(i))=
    \begin{cases}
    A_{k}& \text{for $p=0, i\ge 0$}\\
    A_{-i-1-n}^*& \text{ for $p=n, i\le -n-1$}\\
    0& \text{otherwise}
    \end{cases}
\end{equation}    
\end{proposition}

This proposition and the ampleness of the sequence $\{ \O(i) \}$ implies 
the following corollary:
\begin{cor}\label{van}
Let $X$ be either $\PTh$ or $\Plh$. Then for any sheaf $\F\in coh(X)$ 
and for all
sufficiently large $i\ge 0$ we have
$$
\Hom(\F, \O(i))=0.
$$
\end{cor}
\begin{proof}
By ampleness a sheaf $\F$ can be covered by a finite sum of sheaves 
$\O(k_{j})$. Now the statement follows from the
Proposition, because $\Hom(\O(k_{j}), \O(i))=0$ for all $i< k_{j}$.
\end{proof}
\begin{cor}\label{zero}
Let $X$ be either $\PTh$ or $\Plh$. Then for any sheaf $\F\in coh(X)$
and for all
sufficiently large $i\ge 0$ we have
$$
H^{k}(X, \F(i))=0
$$
for all $k\ge 1$.
\end{cor}
\begin{proof}
The group $H^k(X, \F(i))$ coincides with $\Ext^{k}(\O(-i), \F)$.
Let $k$ be the maximal integer (it exists because the global
dimension is finite) such that for some $\F$ there
exists arbitrarily large $i$ such that $\Ext^{k}(\O(-i), \F)\ne0$.
Assume that $k\ge1$. Choose an epimorphism 
$\mathop{\op}\limits_{j=1}^{s}\O(-k_{j}) \to \F$.
Let $\F_1$ denote its kernel. Then for $i>\max\{k_j\}$
we have $\Ext^{>0}(\O(-i),\mathop{\op}\limits_{j=1}^{s}\O(-k_{j}))=0,$
hence $\Ext^{k}(\O(-i), \F)\ne0$ implies $\Ext^{k+1}(\O(-i), \F)\ne0$.
This contradicts the assumption of the maximality of $k$.
\end{proof}

\subsection{Serre duality and the dualizing sheaf}

A very useful property of commutative 
smooth projective varieties is the existence of the dualizing sheaf.
Recall that a sheaf $\omega$ is called dualizing if
for any $\F\in coh(X)$ there are natural isomorphisms of $\kk$\!--vector spaces
$$
H^i(X, \F)\cong \Ext^{n-i}(\F, \omega)^*,
$$
where $*$ is denotes the $\kk$\!--dual.
The Serre duality theorem asserts the existence of the dualizing sheaf for 
smooth projective
varieties.  In this case the dualizing sheaf is a line bundle
and coincides with the sheaf of differential forms $\Omega_{X}^n$ of top degree.

Since the definition of $\omega$ is given in abstract categorical terms,
it can be extend to the noncommutative case.
More precisely, we will say that $qgr(A)$ satisfies classical Serre duality 
if there is an object $\omega\in qgr(A)$ together with natural isomorphisms
$$
\Ext^{i}(\O,-)\cong \Ext^{n-i}(-, \omega)^*.
$$

Our noncommutative varieties $\PTh$ and $\Plh$ satisfy classical Serre duality,
with dualizing sheaves being $\O_{\PTh}(-4)$ and $\O_{\Plh}(-3),$ respectively.
This follows from the paper \cite{YZ}, where the existence of a dualizing sheaf in $qgr(A)$
has been proved for a general class of algebras which includes all 
noetherian regular algebras. In addition, the authors of \cite{YZ} showed that the 
dualizing sheaf coincides with $\wt{A}(-l),$ where $l$ is the Gorenstein paramenter 
for $A$ (see condition (3) of Definition \ref{reg}).

\subsection{Bundles on noncommutative projective spaces}

To any graded right $A$-module $M$ one can attach a left $A$-module
$M^{\vee}=\Hom_{A}(M, A)$ which is also graded.
Note that under this correspondence the right module $A_{A}(r)$ goes to
the left module ${}_{A}A(-r)$.

It is known that if $A$ is a noetherian regular algebra, then
$\Hom_{A}(-, A)$ is a functor from the category $gr(A)_{R}$ to the category
$gr(A)_{L}$. Moreover, its derived functor $\RHom_{A}^{\cdot}(-, A)$
gives an anti-equivalence between the derived 
categories of $gr(A)_{R}$ and
$gr(A)_{L}$ (see \cite{Y}, \cite{YZ}, \cite{V}).

If we assume that the composition of the functor $\Hom_{A}(-, A)$
with the projection $gr(A)_{L}\lto qgr(A)_{L}$ factors through the projection
$gr(A)_{R}\lto qgr(A)_{R},$ then we obtain a functor from
$qgr(A)_{R}$ to $qgr(A)_{L}$ which is denoted by $\hom( -, \O)$.
This functor is not right exact and has  right derived functors
$\ext^{i}(-, \O),\ i>0,$ from $qgr(A)_{R}$ to $qgr(A)_{L}$.

For a noetherian regular algebra the functor $\hom(-, \O)$ and 
its right derived functors
exist. This follows from the fact that the functors $\Ext_{A}^{i}(-, A)$
send a finite dimensional module to a finite dimensional module (see condition (3) of 
Definition \ref{reg}).

Moreover, in this case the functor $\hom(-, \O)$ can be represented
as the composition of the functor $\Gamma: qgr(A)_{R}\lto gr(A)_{R},$
the functor $\Hom_{A}(-, A): gr(A)_{R}\lto gr(A)_{L},$ and
the projection $\pi: gr(A)_{L}\lto qgr(A)_{L}$.
This can be illustrated by the following commutative diagram:
\begin{equation}\label{comdi}
\begin{array}{ccc}
gr(A)_{R}&  \stackrel{\Hom_{A}(-, A)}{\llongrightarrow} &  gr(A)_{L} \\
\llap{\ss{\pi}}\da \ua\rlap{\ss{\Gamma}} &                &   \da\rlap{\ss{\pi}} \\
qgr(A)_{R} & \stackrel{\hom(-, \O)}{\llongrightarrow} & qgr(A)_{L}
\end{array}
\end{equation}

For a noetherian regular algebra the functor 
$\RHom^{\cdot}_{A}(-, A)$
is an anti-equivalence between the derived categories of $gr(A)_{R}$ and $gr(A)_{L}$
and  takes complexes of finite dimensional modules over $gr(A)_{R}$
to complexes
of finite dimensional modules over $gr(A)_{L}$. This implies that
the functor $\rhom^{\cdot}(-, \O)$ gives an anti-equivalence between the derived categories of
$qgr(A)_{R}$ and $qgr(A)_{L}$. (Note that for derived functors 
$\RHom_{A}(-, A)$ and $\rhom(-, \O)$ there is also a commutative diagram like 
(\ref{comdi})).
 
The functors $\ext^{j}(-, \O)$ can be described more explicitly.
Let $M$ be an $A$\!--bimodule. Regarding it as a right module,
we see that for any $\F\in QGr(A)_{R}$ the groups
$\Ext^i(\F, \wt{M})$ have the structure of left $A$\!--modules.
We can project them to $QGr(A)_{L}$. Thus each bimodule $M$
defines functors from
$QGr(A)_{R}$ to $QGr(A)_{L},$ which will be denoted by
$\pi\Ext^{i}(-, \wt{M})$.

Now, using $\pi\Gamma=id$ and the commutativity of the diagram
(\ref{comdi}) for the derived functors $\Ext^{j}_{A}(-, A)$ and 
$\ext^{j}(-, \O),$ we obtain  isomorphisms
\begin{equation}\label{dualiso}
\ext^{j}(\F, \O)\cong \pi\Ext^{j}_{A}(\Gamma(\F), A)\cong
\pi\Ext^{j}_{gr(A)}(\Gamma(\F), \mathop{\oplus}\limits_{i\ge 0}A(i))
\cong
\pi\Ext^{j}(\F, \mathop{\oplus}\limits_{i\ge 0}\O(i))
\end{equation}
for any sheaf $\F\in qgr(A)_{R}$.
\begin{definition}\label{bund} We call a coherent sheaf $\F\in qgr(A)_{R}$ 
locally free
(or a bundle) if $\ext^{j}(\F, \O)=0$ for any $j\ne 0$.
\end{definition}
\noindent{\sc Remark.}\ In the commutative case this definition is equivalent to
the usual definition of a locally free sheaf.
\begin{definition}
The dual sheaf $\hom(\F, \O)\in qgr(A)_{L}$ will be denoted by $\F^{\vee}$ .
\end{definition}
If $\F\in qgr(A)_{L}$ is a bundle, then the dual sheaf $\F^{\vee}$ 
is a bundle in $qgr(A)_{L},$ because $\rhom^{\cdot}(\F^{\vee}, \O)=\F$ in the derived category, 
and $\ext^{j}(\F^{\vee}, \O)=0$ for $j\ne 0$.

Thus we have a good definition of 
locally free sheaves on $\PTh$ and $\Plh$.
Since the derived functor $\rhom(-, \O)$ gives an anti-equivalence 
between the derived
categories of $qgr(A)_{R}$ and $qgr(A)_{L},$ 
there is an isomorphism:
\begin{equation}\label{dualis}
\Hom(\F, \G)\cong \Hom(\G^{\vee}, \F^{\vee})
\end{equation}
for any two bundles $\F$ and $\G$ on $\PTh$ or $\Plh$.

\section{Bundles on $\Plh$}\label{bundles}

\subsection{Bundles on $\Plh$ with a trivialization on the commutative line}

In this section we study bundles on $\Plh$. By definition,
a bundle is an object $\E\in coh(\Plh)$ satisfying the additional condition 
$\ext^{i}(\E, \O)=0$ for all $i>0$ (see (\ref{bund})).

The noncommutative plane $\Plh$ contains the 
commutative projective line $l\cong \P^1$  given by the equation
$w_{3}=0$. If $M$ is a $\plh$-module, then the quotient module $M/Mw_3$ is a 
$\plh/\langle w_3 \rangle$-module. This gives a functor $coh(\Plh)\to coh(\P^1),$ 
$\F\mapsto\F|_{l}$.
The sheaf $\F|_l$ is referred to as the restriction of $\F$ to the line $l$.
\begin{lemma}
If $\F$ is a bundle, there is an exact sequence:
\begin{equation}\label{ses}
0\lto \F(-1)\stackrel{\cdot w_3}{\lto}\F\lto \F|_{l}\lto 0.
\end{equation}
\end{lemma}
\begin{proof}
To prove this we only need to show that multiplication by 
$w_3$ is a monomorphism. If $\F$ is a bundle, it can be embedded into a direct sum 
$\mathop{\op}\limits_{i=1}^{s} \O(k_{i}),$ because by ampleness the dual
bundle $\F^{\vee}$ is covered by a direct sum of line bundles.
Now, since the morphism $\O(k_i-1)\stackrel{\cdot w_3}{\lto}\O(k_i)$ is mono
for any $i,$ the same is true for the morphism 
$\F(-1)\stackrel{\cdot w_3}{\lto}\F$.
\end{proof}
\begin{lemma}\label{vanish} Let $\E$ be a bundle on $\Plh$ such that its restriction $\E |_{l}$
to the commutative line $l$ is isomorphic to a trivial bundle $\O_{l}^{\op r}$.
Then 
$$
H^{0}(\Plh, \E(-1))=H^{0}(\Plh, \E(-2))=H^2(\Plh, \E(-1))=H^2(\Plh, \E(-2))=0.
$$
\end{lemma}
\begin{proof} 
We have the following exact sequence in the category $coh(\Plh)$:
\begin{equation}\label{esE}
0\lto \E(-2)\lto \E(-1)\lto \E(-1)|_{l}\lto 0.
\end{equation}
Since $\E(-1)|_{l}\cong \O_{l}(-1)^{\op r},$ we have $H^0(\E(-1)|_{l})=0$.

Assume that $\E(-1)$ has a nontrivial section. 
Then $\E(-2)$ has a nontrivial section too. For the same reason 
$\E(-3)$ has a nontrivial section, and so on.
Thus for any $n<0$ the bundle $\E(-n)$ has a nontrivial section.
By (\ref{dualis}) we have isomorphisms:
$$
H^0(\E(-n))\cong \Hom(\O(n), \E)\cong \Hom(\E^{\vee}, \O(-n)).
$$
On the other hand, by Corollary \ref{van} the last group is trivial for $n\gg 0$.
Hence $H^0(\E(-n))=0$ for all $n\gg 0,$ and consequently $H^0(\E(-2))=H^0(\E(-1))=0$.

Further, assume that $H^2(\E(-2))$ is nontrivial. Since $H^1(\E(i)|_{l})=0$
for all $i\ge -1$ we have from the   exact sequence (\ref{ses}) with $\F=\E(i)$
that $H^2(\E(i))$ is nontrivial too for
all $i\ge -1$. But this contradicts Corollary \ref{zero}.
Therefore $H^2(\E(-2))=H^2(\E(-1))=0$. This completes the proof.
\hfill \end{proof}

\subsection{Monads on $\Plh$ and $\PTh$}

As in the commutative case, a non-degenerate monad on $\Plh$ or $\PTh$ 
is a complex over $coh(\Plh)$
$$
0\lto H\ot \O(-1)\stackrel{m}{\lto} K\ot \O \stackrel{n}{\lto} L\ot \O(1)\lto 0
$$ 
for which the map $n$ is an epimorphism and $m$ is a monomorphism. 
(Note that there is another more restrictive definition of a monad, 
according to which the dual map $(m)^*$ has to be an epimorphism, see \cite{OSS}). 
The coherent sheaf 
$$
E=\Ker(n)/\Im(m)
$$
is called the cohomology of a monad. 
A morphism between two monads is a morphism of complexes.
The following lemma is proved in \cite{OSS}(Lemma 4.1.3) in the commutative case,
but the proof is categorical and applies to the noncommutative case as well.
\begin{lemma} Let $X$ be either $\Plh$ or on $\PTh,$
and let $E$ and $E'$ be the cohomology bundles of two
monads
$$
\begin{array}{ll}\label{homon}
M:& 0\lto H\ot \O(-1)\stackrel{m}{\lto} 
K\ot \O \stackrel{n}{\lto} L\ot \O(1)\lto 0,\\
M':& 0\lto H'\ot \O(-1)\stackrel{m'}{\lto} 
K'\ot \O \stackrel{n'}{\lto} L'\ot \O(1)\lto 0
\end{array}
$$
on $X$. Then the natural mapping
$$
\Hom(M, M')\lto \Hom(E, E')
$$
is bijective.
\end{lemma}
The proof is based on the fact that
$$
\Ext^{j}(\O, \O(-1))=\Ext^j(\O(1), \O(-1))=\Ext^j(\O(1), \O)=0
$$
for all $j$
(see \cite{OSS}, Lemma 4.1.3).

\subsection{Non-degeneracy conditions}

In the definition of a monad we require that the map
$n$ be an epimorphism.
In the commutative case this condition must be verified pointwise.
In the noncommutative case the situation is simpler in some sense, because 
the complement of the commutative line $l$ does not have points.
\begin{lemma} If the restriction of a sheaf $\F\in coh(\Plh)$ to the projective
line $l$ is the zero object, then $\F$ is also the zero object.
\end{lemma}
\begin{proof} Let $M$ be a finitely generated graded $\plh$-module 
such that $\F\cong \widetilde{M}$.
Consider an exact sequence:
$$
M\stackrel{\cdot w_3}{\lto} M(1)\lto N\lto 0.
$$
Since $\wt{N}=\F(1)|_{l}=0,$ the module $N$ is finite dimensional. This implies that for 
$i \gg 0$ the map 
$M_{i}\stackrel{\cdot w_3}{\to} M_{i+1}$ is surjective. Moreover, 
these maps are isomorphisms for $i\gg 0,$ because all $M_{i}$
are finite dimensional vector spaces.
Let us identify all $M_{i}$ for $i\gg 0$ with respect to these isomorphisms.
Using the $A$-module structure on $M$, we obtain a representation of the Weyl
algebra $\T( X, Y )/\langle [X, Y]=2\h \rangle$ on the vector space $M_{i}$. 
But it is well known that the Weyl algebra does not have finite 
dimensional representations. Thus $M_{i}=0$ for all $i\gg 0,$ and
$M$ is finite dimensional. Therefore $\F=0.$\hfill\end{proof}

The following corollary is an immediate consequence of the Lemma.
\begin{cor}\label{surj} Let $f:\F\lto \G$ be a morphism in $coh(\Plh)$. Suppose its
restriction $\bar{f}: \F|_{l}\lto \G|_{l}$ is an epimorphism. 
Then $f$ is an epimorphism too.
\end{cor}

\subsection{From the resolution of the diagonal to a monad}

Let $M$ be an $A$\!--bimodule. Regarding it as a left module,
we see that for any $\F\in QGr(A)_{L}$ the groups
$\Ext^i(\F, \wt{M})$ have the structure of right $A$\!--modules.
We can project them to $QGr(A)_{R}$. Thus each bimodule $M$
defines functors $\pi\Ext^{i}(-, \wt{M})$ from
$QGr(A)_{L}$ to $QGr(A)_{R}$.

Let $\E$ be a bundle on $\Plh$ such that its restriction to the 
line $l$
is a trivial bundle.
Let us consider the bundle $\E^{\vee}(1)\in qgr(\plh)_{L}$ and the 
resolution of the diagonal $\CK_{\cdot}(\plh),$ which has only three terms:
$$
\{ 0\lto\plh(-1)\ot\plh(-2)\lto \varOmega^1(1)\ot \plh(-1)\lto \plh\ot \plh \}
\lto \Delta.
$$
The resolution of the diagonal is a complex of bimodules. It induces
a complex $\wt{\CK}_{\cdot}$ over $QGr(\plh)_{L}$:
\begin{equation}\label{resd}
\{0\lto\O(-1)\ot \plh(-2)\lto \Omega^1(1)\ot \plh(-1)\lto \O\ot \plh \}
\lto \wt{\Delta}
\end{equation}
where $\Omega^1$ is a sheaf on $\Plh$ corresponding to the $\plh$\!--module $\varOmega^1$.

As described above, each $A$\!--bimodule $M$ gives the functors
$\pi\Ext^{i}(-, \wt{M})$ from
$QGr(A)_{L}$ to $QGr(A)_{R}$.
In particular, each object of the resolution of the diagonal 
induces such functors.

First we calculate these functors for the object $\wt{\Delta}$.
Note that the object $\wt{\Delta}$ coincides with 
$\mathop{\op}\limits_{i\ge 0}\O(i)$.
Hence by (\ref{dualiso}) we have
$$
\pi\Ext^{j}(\E^{\vee}(1), \wt{\Delta})=0
$$
if $j>0,$ while
$
\pi\Ext^{0}(\E^{\vee}(1), \wt{\Delta})\cong \E(-1).
$

The resolution of the diagonal (\ref{resd}) 
gives us a spectral sequence with the $E_1$ term
$$
E_{1}^{pq}=\pi\Ext^{q}(\E^{\vee}(1), \wt{\CK}_{-p}) \Longrightarrow 
\pi\Ext^{p+q}(\E^{\vee}(1), \wt{\Delta}),
$$
which converges to 
$$
E^{i}_{\infty}=
\begin{cases} 
\E(-1)&\text{if $i=0$}\\
0& \text{otherwise}
\end{cases}
$$

Now we describe all terms $E_{1}^{pq}$ of this spectral sequence.
First we have
$$
\pi\Ext^{j}(\E^{\vee}(1), \O\ot \plh)\cong\Ext^{j}(\E^{\vee}(1), \O)\ot \wt{\plh}
\cong \Ext^{j}(\E^{\vee}(1), \O)
\ot \O
\cong H^{j}(\Plh, \E(-1))\ot \O.
$$
By Lemma \ref{vanish}, these groups are trivial for $j\ne 1$.
For the same reason we have
$$
\pi\Ext^{j}(\E^{\vee}(1), \O(-1)\ot \plh(-2))=H^{j}(\Plh, \E(-2))\ot \O(-2)=0
$$
for $j\ne 1$ and
$$ 
\pi\Ext^{1}(\E^{\vee}(1), \O(-1)\ot \plh(-2))\cong H^{1}(\Plh, \E(-2))\ot \O(-2).
$$
Now let us consider the functors which are associated with the object 
$\Omega^1(1)\ot \plh(-1)$. We have
$$
\pi\Ext^{j}(\E^{\vee}(1), \Omega^1(1)\ot \plh(-1))\cong \Ext^{j}(\E^{\vee}, 
\Omega^1)
\ot \O(-1).
$$
It follows from the Koszul complex that the sheaf $\Omega^1$ 
can be included in two exact sequences:
$$
\begin{array}{c}
0 \lto \Omega^1 \lto \O(-1)\ot \plh_1 \lto \O \lto 0,\\
0 \lto \O(-3)\lto \O(-2)\ot (\plh_1)^* \lto \Omega^1 \lto 0.
\end{array}
$$
Applying the functor $\Hom(\E^{\vee}, -)$ to the first sequence and taking into account 
that $\Hom(\E^{\vee}, \O(-1))=0,$ we obtain
$\Hom(\E^{\vee}, \Omega^1)=0$.
Similarly,  we deduce from the second sequence that 
$\Ext^{2}(\E^{\vee}, \Omega^1)=0,$ because 
$\Ext^{2}(\E^{\vee}, \O(-2))=0$.
This implies that the object
$\pi\Ext^{j}(\E^{\vee}(1), \Omega^1(1)\ot \plh(-1))$ is trivial for all 
$j\ne 1$.

Thus our spectral sequence is nothing more than the complex
$$
\pi\Ext^{1}(\E^{\vee}(1), \wt{\CK}_{2})\lto
\pi\Ext^{1}(\E^{\vee}(1), \wt{\CK}_1)\lto
\pi\Ext^{1}(\E^{\vee}(1), \wt{\CK}_0),
$$
which is isomorphic to the complex
$$
H^{1}(\Plh, \E(-2))\ot \O(-2)\lto\Ext^{1}(\E^{\vee}, \Omega^1)
\ot \O(-1)\lto H^{1}(\Plh, \E(-1))\ot \O.
$$
It has only one cohomology which coincides with
$\E(-1)$.
\begin{theorem} Let $\E$ be a bundle on $\Plh$ such that its restriction
to the commutative line $l$ is isomorphic 
to the trivial bundle
$\O_{l}^{\op r}$. Then $\E$ is the cohomology of a monad 
$$
0\lto H\ot \O(-1)\stackrel{m}{\lto} K\ot \O \stackrel{n}{\lto} L\ot \O(1)\lto 0
$$ 
with $H=H^{1}(\Plh, \E(-2)), \ L=H^1(\Plh, \E(-1)),$ and such monad is unique up 
to an isomorphism. Moreover, in this case the vector spaces $H$ and $L$ have the
same dimension.
\end{theorem}
\begin{proof} The existence of such a monad was proved above. 
The uniqueness follows
from Lemma \ref{homon}. The equality of dimensions of $H$ and $L$ follows
immediately from the exact sequence (\ref{esE}).
\end{proof}

\subsection{Barth description of monads}

Now following Barth \cite{Ba}, we give a description of the moduli space
of vector bundles on $\Plh$ trivial on the line $l$ in terms of linear algebra
(see also \cite{Do}).

Denote by $\M_{\h}(r, 0, \ctwo)$ the moduli space of bundles
on the noncommutative $\Plh$ trivial on the line $l$
and  with a fixed trivialization there (i.e. with a fixed isomorphism
$\E|_{l}\cong \O_{l}^{\op r}$). Let $\dim H^{1}(\Plh, \E(-1))=\ctwo$.
As in the commutative case, the numbers $r, 0, \ctwo$ can be regarded as the rank, 
first Chern class, and second Chern class of $\E,$ respectively.

The following theorem gives a description of this moduli space
which is similar to the description given by Barth in the commutative case.
\begin{theorem}Let 
$\{(\bnul, \bone;\mui, \nui)\}$ be the set of quadruples of matrices
$\bnul, \bone\in M_{\ctwo\times \ctwo}(\C),\ 
\mui\in M_{r\times \ctwo}(\C),\ \nui \in M_{\ctwo\times r}(\C),$ 
which  satisfy the condition
$$
[\bnul, \bone]+\nui\mui + 2\h\cdot 1_{\ctwo\times \ctwo}=0.
$$
Then the space $\M_{\h}(r, 0, \ctwo)$ is the quotient of this set 
with respect to the following free action of $\GL(\ctwo, \C)$:
$$
b_{i}\mapsto g b_{i} g^{-1},\quad
\mui\mapsto \mui g^{-1},\quad
\nui\mapsto g\nui,\quad\text{where $g\in \GL(\ctwo, \C)$}.
$$
\end{theorem}
\begin{proof}
Let $\E$ be a bundle on $\Plh$ trivial on the line $l$.
We showed above that any such bundle comes from
a monad unique up to an isomorphism.
Conversely, suppose we have a monad
\begin{equation}\label{mond}
0\lto H\ot \O(-1)\stackrel{m}{\lto} K\ot \O \stackrel{n}{\lto} L\ot \O(1)\lto 0
\end{equation}
with $\dim H=\dim L=\ctwo$ such that its restriction to the line $l$ is a monad with
the cohomology $\O_{l}^{\oplus r}$. Then the cohomology of this monad is a bundle on
$\Plh$ which belongs to $\M_\h(r, 0, \ctwo)$.
Indeed, the cohomologies of the dual complex 
$$
0\lto  \O(-1)\ot L^*\stackrel{n^*}{\lto}  \O\ot K^* \stackrel{m^*}{\lto} \O(1)\ot H^*\lto 0
$$
coincide with $\hom(\E, \O)$ and 
$\ext^{1}(\E, \O)$.
Hence, to prove that $\E$ is a bundle, it is sufficient to show that
the dual complex is a monad too, i.e. that the map $m^*$ is an epimorphism.
The restriction of the dual complex to $l$ is a monad which is dual to the restriction of 
the monad (\ref{mond}) to $l$. Hence the restriction of $m^*$ on $l$ is an epimorphism.
Then, by Lemma \ref{surj}, $m^*$ is an epimorphism as well.
Thus to describe the moduli space $\M_{\h}(r, 0, \ctwo)$ we have to decsribe
the space of all monads (\ref{mond}) modulo isomorphisms preserving trivialization on $l$.

Consider a monad 
$$
0\lto H\ot \O(-1)\stackrel{m}{\lto} K\ot \O \stackrel{n}{\lto} L\ot \O(1)\lto 0
$$ 
with $\dim H= \dim L= \ctwo$ and $\dim K=2\ctwo+r$. Denote by $\E$ its cohomology bundle.

The maps $m$ and $n$ can be regarded as elements of $H^*\ot K\ot W$ and $K^*\ot L\ot W,$
respectively, where $W=H^{0}(\Plh, \O(1))$ is the vector space spanned by $w_1, w_2, w_3$.
The maps $m$ and $n$ can be written as
$$
m_{1}w_1 + m_2 w_2 + m_3w_3, \quad n_{1}w_1 + n_2 w_2 + n_3w_3,
$$
where $m_{i}: H\to K$ and $n_{i}: K\to L$ are constant linear maps.

Let us restrict the monad to the line $l$. The monadic condition $nm=0$
implies now:
$$
n_1 m_2 + n_2 m_1 =0,\quad  n_1 m_1=0,\quad n_2 m_2=0.
$$
Moreover, since the restriction of $\E$ to $l$ is trivial, the composition
$n_1 m_2$ is an isomorphism (see \cite{OSS}, Lemma 4.2.3).
We can  choose bases  for $H, K, L$ so that $n_1m_2=1_{\ctwo\times \ctwo}$ 
(the identity matrix) and
$$
m_{1}=
\begin{pmatrix} 1_{\ctwo\times \ctwo}\\ 0_{\ctwo\times \ctwo} \\ 
0_{r\times \ctwo}\end{pmatrix},\ \ 
m_2=
\begin{pmatrix} 0_{\ctwo\times \ctwo}\\ 1_{\ctwo\times \ctwo} \\ 
0_{r\times \ctwo}\end{pmatrix},\ \ 
n_1=
\begin{pmatrix} 0_{\ctwo\times \ctwo}& 1_{\ctwo\times \ctwo} & 0_{\ctwo\times r}\end{pmatrix},\ \ 
n_2=
\begin{pmatrix} -1_{\ctwo\times \ctwo}& 0_{\ctwo\times \ctwo} & 0_{\ctwo\times r}\end{pmatrix}.
$$
Using the equations $n_3 m_1+ n_1m_3=0$ and 
$ n_3 m_2+ n_2m_3=0$ we can write:
$$
m_{3}=
\begin{pmatrix} \bnul\\ \bone \\ \mui\end{pmatrix},\quad
n_3=
\begin{pmatrix} -\bone& \bnul & \nui\end{pmatrix}.
$$
Now the monadic condition $nm=0$ can be written as:
$$
(n_3 m_3 )\cdot w_{3}^2 + 1_{\ctwo\times \ctwo}\cdot[w_1, w_2]=0.
$$
Therefore we obtain the following matrix equation:
$$
[\bnul, \bone]+\nui\mui + 2\h \cdot 1_{\ctwo\times \ctwo}=0.
$$
Note that the last $r$ basis vectors of $K$ give us a trivialization of the
restriction of $\E$ to the line $l$. It is easy to check that any isomorphism of a monad 
which preserves trivialization on $l$ and the choice of the bases of $H, K, L$
made above has the form 
$$
b_{i}\mapsto g b_{i} g^{-1},\quad
\mui\mapsto \mui g^{-1},\quad
\nui\mapsto g\nui,\quad\text{where $g\in \GL(\ctwo, \C)$}.
$$
This proves the theorem.
\end{proof}

\section{The noncommutative variety $\PTh$ as a twistor space}\label{pthree}

\subsection{Real structures}\label{ptwistor}

A $*$\!--algebra is, by definition, an algebra over $\C$ with an anti-linear anti-homomorphism
$*$ satisfying $*^{2}=id$. A $*$\!--structure on a (graded) algebra  is
regarded as a real structure on the corresponding (projective) noncommutative variety.

Let us introduce real structures on the complex  varieties $\C^{4}_{\h}$ and $\Q^{4}_{\h}$ 
defined in section~\ref{varieties}. 
Assume that in (\ref{cfour}), (\ref{quadrel}) the skew-symmetric matrix $\theta$ is purely 
imaginary and $\h$ is real. Then there is a unique $*$\!--structure on the 
algebra $A(\C_{\h}^{4})$ such that  $x_{i}^{*}= x_i$. We denote the corresponding 
noncommutative variety by $\R^{4}_{\h}$.

Assume in addition that the symmetric matrix
$G$ in (\ref{quadrel}) is real and positive definite. 
There is a unique $*$\!--structure on the algebra $Q_{\h}$
such that  $X_{i}^{*}= X_i,$$D^*=D,$ and $T^*=T$.
The corresponding noncommutative real variety will be called the noncommutative sphere
and denoted by $\S^{4}_{\h}$. The embedding of $\C^{4}_{\h}$ into $\Q_{\h}^{4}$
induces an embedding $\R^{4}_{\h}\hookrightarrow \S^{4}_{\h}$. Recall that the complement of
$\C^{4}_{\h}$ in $\Q_{\h}^{4}$ is a commutative quadratic cone 
$\sum\limits_{kl} G^{kl} X_{k}X_{l}=0$
which has only one real point. Thus $\S^{4}_{\h}$ can be regarded as a one-point 
compactification of~$\R^{4}_{\h}$.

By a linear change of basis one can bring the pair $(G,\theta)$ to the standard form
\begin{equation}\label{standard}
G=\begin{pmatrix} 
1 & 0 & 0 & 0 \\ 
0 & 1 & 0 & 0 \\
0 & 0 & 1 & 0 \\
0 & 0 & 0 & 1 
\end{pmatrix}, 
\qquad
\theta=\sqrt {-1}
\begin{pmatrix}
0 & a & 0 & 0\\
-a & 0 & 0 & 0\\
0 & 0 & 0 & b\\
0 & 0 & -b & 0
\end{pmatrix}.
\end{equation}
Furthermore, since $\h$ and $\theta$ enter only in the combination $\h\cdot \theta,$
and we asssume that $a+b\neq 0,$ we can set $a+b=1$ without loss of generality.

\subsection{Realification of $\PTh$}

Recall that the noncommutative projective space $\PTh$ corresponds to
the algebra $\pth$ with generators $z_i,i=1,2,3,4,$ and relations (\ref{rel}). 
Consider an algebra $\prh$ with generators $z_i,\ \bz_i, i=1,2,3,4,$
and relations
\begin{equation}\label{zbzrel}
\arraycolsep=0pt
\begin{array}{l}
\begin{array}{lll}
{}[z_1,z_2]=\hphantom{-}2\h(a+b)z_3z_4, \quad			
& [z_1,\bz_1]=2\h b z_3\bz_3-2\h az_4\bz_4,\quad & [z_1,\bz_2]=0, \\
{}[\bz_1,\bz_2]=-2\h(a+b)\bz_3\bz_4,\quad	
& [z_2,\bz_2]=2\h a z_3\bz_3-2\h b z_4\bz_4,\quad & [z_2,\bz_1]=0, 
\end{array}\\
{}[z_i,z_j]=[z_i,\bz_j]=[\bz_i,z_j]=[\bz_i,\bz_j]=0\quad\text{for all}
\quad i=3,4; j=1,2,3,4
\end{array}
\end{equation}

There is a unique $*$\!--structure on this algebra 
such that  $z_{i}^{*}= \bz_i,$$\bz_{i}^{*}= z_i$.
We denote the corresponding real variety $\PP^3_\h(\R)$.
This variety can be considered a realification of $\PTh$.

{\noindent\sc Remark.}
In contrast to the commutative situation, a noncommutative complex variety
in general has many different realifications. We have an
ambiguity in the choice of relations involving both $z_i$ and $\bz_j$.
The realification (\ref{zbzrel}) is distinguished 
by the fact that it is the twistor space of the noncommutative
sphere $\S^{4}_{\h},$ as explained below.

In the commutative case there is a map from $\PP^{3}(\R)$ to the sphere
$\S^4$ which is a $\PP^1$ fibration. The corresponding $\PP^1$ bundle is the projectivization
of a spinor bundle on $\S^4$. This map is known as the Penrose map. 
In the noncommutative case we have a similar picture. The analogue of the Penrose map 
is a map $\Pi: \PTh(\R)\lto \S^{4}_{\h}$ which is associated with the homomorphism of 
$*$\!-algebras  $Q_{\h}\lto \prh$:
$$
\begin{array}{llcllll}
X_1 & \mapsto & \displaystyle -\frac{\sqrt{-1}}2 & (z_1\bz_4-\bz_1z_4-\bz_2z_3+z_2\bz_3), \quad
D   & \mapsto & \displaystyle -\frac12\ (z_1\bz_1+\bz_1z_1+z_2\bz_2+\bz_2 z_2), \medskip\\
X_2 & \mapsto & \displaystyle  \frac12 & (z_1\bz_4+\bz_1z_4-\bz_2z_3-z_2\bz_3),\quad 
T   & \mapsto & -\ \ \ (z_3\bz_3+z_4\bz_4),\medskip\\
X_3 & \mapsto & \displaystyle -\frac{\sqrt{-1}}2 & (\bz_1z_3-z_1\bz_3+z_2\bz_4-\bz_2z_4), \medskip\\
X_4 & \mapsto & \displaystyle \frac12 & (z_1\bz_3+\bz_1z_3+\bz_2z_4+z_2\bz_4). \\
\end{array}
$$
Note that for $\h=0$ we obtain the homomorphism of commutative algebras which corresponds to
the usual Penrose map. 
This means that $\PTh(\R)$ is the twistor space of $\S^4_{\h}$.

The variety $\PTh(\R)$ is a twistor space in yet another sense.
For the commutative $\RR^4$ the complex structures compatible with
the symmetric bilinear form $G$ and orientation are parametrized by points of a $\PP^1$. 
This remains true in the noncommutative case. 
A complex structure (resp. orientation) on $\RR^4_{\h}$ is defined as a complex structure 
(resp. orientation) on the real vector space $U$ spanned by $x_1,\ldots,x_4$. We will choose an
orientation on $U$ and require that the complex structure be compatible with it. 
All such complex structures are parametrized by points of a $\PP^1$.
 
Recall now that $\PTh$ is a pencil of noncommutative projective planes passing through
the commutative line. Let us pick any one of them. The realification of $\PTh$ defined above 
induces a realification of the noncommutative projective plane. It is easy to see that the 
complement of the commutative line $w_3=\bar{w}_3=0$ in the realified projective plane is 
isomorphic to $\RR^4_{\h}$. Furthermore, the complement carries a natural complex structure 
defined by 
$$
w_3^{-1}w_i\mapsto\sqrt {-1}\ w_3^{-1}w_i,\qquad\qquad 
\bar{w}_3^{-1}\bar{w}_i\mapsto-\sqrt {-1}\ \bar{w}_3^{-1}\bar{w}_i,\qquad i=1,2.
$$
The Penrose map induces an identification between the complement and $\RR^4_{\h}\subset \S^4_{\h},$
and therefore induces a complex structure on the latter. Varying the noncommutative
projective plane, one obtains all possible complex structures on~$\RR^4_{\h}$ compatible
with a particular orientation.
This is completely analogous to the commutative case.

\subsection{Connection between sheaves on commutative and noncommutative planes}\label{ABC}

In this subsection we are going to connect the moduli space $\M_{\h}(r, 0, \ctwo)$ 
of bundles  on $\Plh$ with a trivialization on the line $l$
with the moduli space $\M(r, 0, \ctwo)$
of torsion free sheaves on the commutative $\Pl$ with a trivialization on
a fixed line. The bridge between bundles on $\Plh$ and torsion free sheaves on $\Pl$
is provided by the twistor variety $\PTh$. This gives a geometrical interpretation
of Nakajima's results (the description of the moduli space $\M(r, 0, \ctwo)$
by the deformed ADHM data 
\cite{Na, Na2}). We will construct a hyperk\"ahler manifold ${\mathcal M}$
parametrizing certain complexes
on $\PTh$ which is isomorphic to $\M(r, 0, \ctwo)$ 
(which is also a hyperk\"ahler manifold \cite{Na}).
The isomorphism is given by the restriction of  complexes  to one of the commutative $\Pl$\!'s.
On the other hand, the restriction of complexes to a noncommutative plane $\Plh$  yields an 
isomorphism between ${\mathcal M}$ with a particular choice of complex structure and the moduli space 
$\M_{\h}(r, 0, \ctwo)$. Thus $\M_{\h}(r, 0, \ctwo)$ can be obtained from $\M(r, 0, \ctwo)$
by a rotation of complex structure.

Consider complexes $\Ko$ on $\PTh$ of the form
\begin{equation}\label{comp}
0\lto H\ot \O(-1)\stackrel{M}{\lto} K\ot \O \stackrel{N}{\lto} L\ot \O(1)\lto 0
\end{equation}
with $\dim H=\dim L=\ctwo,$ $\dim K=2\ctwo+r,$ which satisfies the condition that its restriction
to the line $l$ has only one cohomology which is a trivial bundle (with a fixed trivialization).
This condition implies that $M$ is a monomorphism. Note that
$N$ is not an  epimorphism in general, so (\ref{comp}) is not a monad.
But the restriction of the complex (\ref{comp}) to any noncommutative plane
is a monad by Corollary~\ref{surj}. 
Thus $N$ can fail to be surjective only on the commutative planes 
$z_{3}=0$ and $z_4=0$.

Now we introduce a real structure on $\PTh$ (this is different from the real structure on
the realification of $\PTh$ defined above). 
Assume  that $\h$ is a real number.  Consider 
an anti-linear anti-homomorphism
$\wb{\J}$ of  $\pth$ defined by
$$
\wb{\J}(z_1)= z_2,\quad \wb{\J}(z_2)=-z_1,\quad \wb{\J}(z_3)=z_4,\quad \wb{\J}(z_4)=-z_3,\quad 
\wb{\J}(\lambda)=\wb{\lambda},\quad \lambda\in \C.
$$
Thus $\wb{\J}$ is a homomorphism of $\R$\!--algebras from $\pth$ to the opposite algebra 
$\pth^{op}$. (The notation $\wb{\J}$ is used by analogy with the commutative case, where this
anti-homomorphism is a composition of a complex structure $J$ with
complex conjugation \cite{Do}.)

The anti-homomorphism $\wb{\J}$ induces a functor $\wb{\J}^*$ from $qgr(\pth)_{R}$ to
$qgr({\pth}^{op})_{R}$. The latter category is naturally identified with the category
$qgr(\pth)_{L}$. Using this identification we can consider 
the composition of $\wb{\J}^*$ with the dualization functor $\hom(-, \O)$ as a functor 
from $qgr(\pth)_R$ to itself. For any bundle $\E$ we denote
by $\wb{\J}^*(\E)^{\vee}$ its image under this functor. The functor can be 
extended to complexes of bundles. It takes the complex
$\Ko$ (\ref{comp}) to the complex $\wb{\J}^*(\Ko)^{\vee}$ 
$$
0\lto \wb{L}^{*}\ot \O(-1)\stackrel{\wb{\J}^*(N)^{\vee}}{\lto} 
\wb{K}^{*}\ot \O \stackrel{\wb{\J}^*(M)^{\vee}}{\lto} \wb{H}^{*}\ot \O(1)\lto 0.
$$

Let us consider complexes $\Ko$ on $\PTh$ with an isomorphism 
\begin{equation}\label{realstructure}
\wb{\J}^*(\Ko)^{\vee}\cong \Ko
\end{equation} 
and trivialization on the line $l$.
Then the space $K$ acquires a hermitian metric and $L$ becomes isomorphic to $\wb{H}^*$.
The reasoning of section~\ref{bundles} shows that we can represent the maps $M$ and $N$ as
$$
M_1 z_1 + M_2 z_2 + M_3 z_3 + M_4 z_4 , \quad N_1 z_1 + N_2 z_2 + N_3 z_3 + N_4 z_4,
$$
where $M_{i}$ and $N_{i}$ are constant maps.
By a suitable choice of bases we can put these maps into the form
\begin{equation}\label{form1}
M_{1}=
\begin{pmatrix} 1\\ 0\\ 0\end{pmatrix},\quad
M_2=
\begin{pmatrix} 0\\ 1 \\ 0\end{pmatrix},\quad
M_{3}=
\begin{pmatrix} \Bnul\\ \Bone\\ \Mu\end{pmatrix},\quad
M_{4}=
\begin{pmatrix} \Bnul^{'}\\ \Bone^{'}\\ \Mu'\end{pmatrix},
\end{equation}
$$
N_1=
\begin{pmatrix} 0& 1 & 0 \end{pmatrix},\quad
N_2=
\begin{pmatrix} -1& 0 & 0\end{pmatrix},\quad
N_{3}=
\begin{pmatrix} -\Bone & \Bnul & \Nu\end{pmatrix},\quad
N_{4}=
\begin{pmatrix} -\Bone^{'} & \Bnul^{'} & \Nu^{'}\end{pmatrix}.
$$
Using the reality conditions $\wb{\J}^*(N)^{\vee}= M$ and $\wb{\J}^*(M)^{\vee}=-N$ we find that
\begin{equation}\label{form2}
\Bnul^{'}=-\Bone^{\dag},\quad \Bone^{'}=\Bnul^{\dag},\quad
\Mu^{'}=\Nu^{\dag},\quad \Nu^{'}= -\Mu^{\dag}.
\end{equation}
Finally the condition $NM=0$ gives 
$$
\begin{array}{ll}
a)& \mu_{c}=[\Bnul, \Bone]+ \Nu\Mu=0,\\
b)& \mu_{r}=[\Bnul, \Bnul^{\dag}]+[\Bone, \Bone^{\dag}]+ 
\Nu\Nu^{\dag} -\Mu^{\dag}\Mu=- 2\h\cdot {1}_{\ctwo\times\ctwo}.
\end{array}
$$
These  matrix equations are invariant under the following action of $U(k)$:
\begin{equation}\label{unaction}
B_i\mapsto g B_i g^{-1},\qquad \Nu\mapsto g\Nu, \qquad \Mu\mapsto \Mu g^{-1},\quad 
\text{where $g\in U(k).$}
\end{equation}

Denote by ${\mathbf M}$ the vector space of complex matrices 
$(\Bnul , \Bone, \Nu, \Mu)$. It has a structure of a quaternionic vector space defined by
$$
(\Bnul, \Bone, \Nu, \Mu) \mapsto (-\Bone^{\dag}, \Bnul^{\dag}, -\Mu^{\dag},
\Nu^{\dag}),
$$
and, moreover, it is a flat hyperk\"ahler manifold (see \cite{Na}).
The map $\mu=(\mu_r , \mu_c)$ is a hyperk\"ahler moment map for the action
of $U(k)$ defined in~(\ref{unaction}) (see \cite{HKLR}).
Since the action of $U(k)$ on $\mu_c^{-1}(0)\cap \mu_r^{-1}(-2\h\cdot 1)$
is free, the quotient ${\mathcal M}=\mu_c^{-1}(0)\cap \mu_r^{-1}(-2\h\cdot 1)/U(k)$ 
is a smooth hyperk\"ahler manifold. This manifold parametrizes  complexes (\ref{comp}) with 
a real structure~(\ref{realstructure}) and a trivialization on the line $l$.

On the other hand, it was proved in \cite{Na, Na2} that the moduli space $\M(r, 0, \ctwo)$ of torsion 
free sheaves on the commutative $\Pl$ with a trivialization on a fixed line can be identified
with ${\mathcal M}$.

This identification can be described geometrically as follows.
Let us assume that $\h$ is positive. It can be  checked that
in this case the map $N$ can fail to be surjective only on the plane
$z_4 =0$. We can restrict the complex (\ref{comp})
to the commutative plane $z_3=0$. The restriction is
a monad and its cohomology sheaf is a torsion free sheaf. It is easy to
see that this yields a complex isomorphism from ${\mathcal M}$ to $\M(r, 0, p)$.

The restriction of the complex (\ref{comp}) to a noncommutative plane is a monad as well.
This yields 
a map from ${\mathcal M}$ to
the moduli space $\M_{\h}(r, 0, \ctwo)$ of bundles on the noncommutative plane.
Let us show that this map is an isomorphism.
To this end we note that on the level of the linear algebra data this map sends
a quadruple $(\Bnul, \Bone, \Nu, \Mu)$ to the quadruple $(\bnul, \bone, \nui, \mui)$
with
$$
\begin{array}{l}
\bnul=\Bnul -\Bone^{\dag},\quad \bone=\Bone + \Bnul^{\dag},\quad
\nui=\Nu -\Mu^{\dag},\quad
\mui=\Mu + \Nu^{\dag}.
\end{array}
$$
Further, note that the equations $\mu_c=0,$ $\mu_r=-2\h\cdot 1$ are equivalent
to the equation $[\bnul,\bone]+\nui\cdot\mui+2\h \cdot 1=0$ and the vanishing of
the moment map for the action of the group $U(\ctwo)$ on the space
of quadruples $(\bnul, \bone, \nui, \mui)$. Now it follows
from the theorem of Kempf and Ness (\cite{Na}, \cite{KN}) that 
the map ${\mathcal M}\to\M_{\h}(r, 0, \ctwo)$ is a diffeomorphism.
It becomes a complex isomorphism if we replace the natural 
complex structure of the space ${\mathcal M}$ with another one within
the $\PP^1$ of complex structures on ${\mathcal M}$.

Thus we have
\begin{theorem}
The moduli space $\M_{\h}(r, 0, \ctwo)$ is a smooth hyperk\"ahler manifold
of real dimension $4r\ctwo,$ and as a hyperk\"ahler manifold
it is isomorphic to the moduli space
$\M(r, 0, \ctwo)$ of torsion free sheaves on the commutative $\Pl$ with a trivialization on 
a fixed line. As a complex manifold $\M_{\h}(r, 0, \ctwo)$ is obtained from $\M(r, 0, \ctwo)$
by a rotation of the complex structure.
\end{theorem}

The above discussion shows that there are natural bijections between 

$A'.$ Bundles on $\Plh$ with a trivialization on the commutative line $l$ and $c_2=k.$ 

$B'.$ Solutions of the equations $\mu_c=0,\ \mu_r=-2\h \cdot 1$ modulo the action
of $U(k).$

$C'.$ Complexes of sheaves on $\PTh$ of the form (\ref{comp}) with a trivialization on the 
commutative line $l$ satisfying the reality condition (\ref{realstructure}). 

One can show that for $r>1$ a generic complex (\ref{comp}) is a monad and its cohomology
is a bundle $\E$ on $\PTh$ such that 
\begin{equation} \label{good}
H^{1}(\PTh, \E(-2))=0, \quad \wb{\J}^{*}(\E)^{\vee}\cong \E.
\end{equation}
Moreover, it can be shown that any bundle $\E$ satisfying the conditions (\ref{good}) can be
represented as a cohomology of a monad of the form (\ref{comp}).

\section{Noncommutative twistor transform}\label{twistor}

\subsection{Review of the twistor transform}

In the commutative case the ADHM construction of instantons has the
following geometric interpretation. Consider the double fibration
\begin{equation}\label{ctwistors}
\begin{CD}
\Gr(2;4) @<p<< \Fl(1,2;4) @>q>> \PT,
\end{CD}
\end{equation}
where $\Gr(2;4)$ is the Grassmannian and $\Fl(1,2;4)$ is the partial flag 
variety. The Grassmannian $\Gr(2;4)$ has a real structure with $\S^4$ as the
set of real points. For any bundle $\E$ on $\PP^3$ its twistor transform
is defined as a sheaf $p_*q^*\E$ on $\Gr(2;4)$. Given ADHM data 
we have a monad on $\PP^3$ whose cohomology is a bundle. It can be shown
that the restriction of its twistor transform to the sphere $\S^4$ coincides 
with the instanton bundle corresponding to these ADHM data.
The instanton connection can also be reconstructed from the bundle on 
$\PP^3$ (see \cite{Atiyah,Manin} for details).

In this section we show that one can consider the 
noncommutative quadric introduced in section~\ref{varieties} as 
a noncommutative Grassmannian $\Gr(2;4)$. We also construct 
a noncommutative flag variety $\Fl(1,2;4)$ and projections $p,$ $q$ giving
a noncommutative analogue of the twistor diagram~(\ref{ctwistors}). 
The twistor transform can be defined in the same way as above.
It produces a bundle on the noncommutative sphere from the deformed ADHM data.
We show that this bundle is precisely the kernel of the map ${\mathcal D}$
defined in section~\ref{summary}.

It should also be possible to construct the instanton connection on the noncommutative
$\RR^4$ from the complex of sheaves on $\PTh$. To do this, one needs to
develop differential geometry of noncommutative affine and projective varieties. We go some
way in this direction by defining differential forms and spinors. 

Since the goal of this section is mainly illustrative, we limit ourselves to stating 
the results. An interested reader should be able to fill in the proofs.

\subsection{Tensor categories}

A good way to construct noncommutative varieties with properties
similar to those of commutative varieties is to start with a tensor category (see 
\cite{Ma2,Lyu}).
Let $\TT$ be an abelian tensor category. Consider a tensor functor $\Phi:\TT\to\Vect$ to the 
abelian tensor category of vector spaces compatible with the associativity constraint
but not compatible with the commutativity constraint. If $A$ is a commutative
algebra in the tensor category $\TT,$ then $\Phi(A)$ is a noncommutative algebra
in the tensor category $\Vect$. If $M\in\TT$ is a right $A$-module, then
$\Phi(M)$ is a right $\Phi(A)$-module. Any right $A$\!--module (in the category $\TT$) 
has a natural structure of a left $A$\!--module (and hence an $A$\!--bimodule). Thus 
any right $\Phi(A)$\!--module of the form $\Phi(M)$ has a natural structure of a 
$\Phi(A)$\!--bimodule.

Consider the category $\Comm_\TT$ of all finitely generated (graded) commutative
algebras in the tensor category $\TT$. Then under $\Phi$ the category $\Comm_\TT$ is mapped
to a subcategory of the category of finitely generated (graded) algebras. This subcategory
enjoys many properties of the category of commutative (graded) algebras. For example, for all
$A,B\in \Comm_\TT$ there is a natural algebra structure on $\Phi(A)\ot \Phi(B)$
coming from the algebra structure on $A\ot B.$
The corresponding subcategory in the category of noncommutative affine
(resp. projective) varieties shares a lot of properties with the category of commutative 
varieties. For example,
if $X$ and $Y$ are varieties in this category, then using the tensor
product of the corresponding algebras one can define
the ``Carthesian'' product $X\times Y$. More generally, given a pair of morphisms
$X\to Z$ and $Y\to Z$ one can define the fiber product $X\times_ZY$. Further,
starting from the module of differential forms of $A$ one can construct 
the sheaf of differential forms on the corresponding noncommutative variety.

The category $qgr(\Phi(A))$ has a nice subcategory which consists of
modules of the form $\Phi(M),$ where $M\in \TT$ is an $A$\!--module. 
To any object $\Phi(M)$ of this subcategory
one can associate its symmetric and exterior powers. The symmetric
powers of $\Phi(M)$ form a noncommutative graded algebra. This enables one to define
the projectivization of the sheaf corresponding to the module $\Phi(M).$

\subsection{Yang-Baxter operators}

One way to construct an abelian tensor category $\TT$ with a functor $\Phi:\TT\to\Vect$
is to consider a Yang-Baxter operator (see \cite{Ma2}, \cite{Lyu}).

A Yang-Baxter operator on a vector space $V$ is an 
operator
$R:V\ot V\to V\ot V,$ such that 
\begin{equation}\label{YB}
\arraycolsep=0pt
R^2=\id_{V\ot V},\quad
(R\ot\id_V)(\id_V\ot R)(R\ot\id_V)=
(\id_V\ot R)(R\ot\id_V)(\id_V\ot R).
\end{equation}
A Yang-Baxter operator induces an action of the permutation group  
$\FS_n$ on the tensor power $V^{\ot n},$ where the transposition
$(i,i+1)\in\FS_n$ acts as the operator
$$
R_{i,i+1} = \id_{V^{\ot(i-1)}} \ot R \ot \id_{V^{\ot(n-i-1)}}:
V^{\ot n} \to V^{\ot n}.
$$
Equations (\ref{YB}) ensure that operators $R_{i,i+1}$ satisfy
the relations between the transpositions $(i,i+1)$ in the group $\FS_n$.

If $R$ is a Yang-Baxter operator on a vector space $V,$
then the dual operator $R^\vee:V^*\ot V^*\to V^*\ot V^*$ is also
a Yang-Baxter operator.

Given a Yang-Baxter operator $R:V\ot V\to V\ot V,$ one can construct
an abelian tensor category $\TT_R$ and a functor $\Phi_R: \TT_R\to\Vect$ such
that $V$ is a $\Phi_R$\!--image of some object of $\TT_R$, and the commutativity morphism 
in the category $\TT_R$ is mapped by
$\Phi_R$ to $R$~\cite{Lyu}. As mentioned above, given any two objects $A,B$ of the category
$\Comm_{\TT_R},$ one has a natural algebra structure on the vector space 
$\Phi(A)\ot\Phi(B)$.
This algebra will be denoted $\Phi(A)\otR\Phi(B)$ and called the $R$-tensor 
product of $\Phi(A)$ and $\Phi(B)$.

It is well known that there is a one-to-one correspondence between
irreducible representations of the group $\FS_n$ and
partitions of $n$ (Young diagrams). Under this correspondence the trivial partition $(n)$
corresponds to the sign representation, while the maximal partition
$(\underbrace{1,1,\dots,1}_{\text{$n$ times}})$ corresponds to the 
identity representation. Given a partition $(k_1,\dots,k_r)$ of~$n$
$(k_1\ge k_2\ge\dots\ge k_r)$ we denote by $(k_1,\dots,k_r)$ the
corresponding irreducible representation and by 
$\Sigma_R^{(k_1,\dots,k_r)} V$ (resp.\ $\Sigma_R^{(k_1,\dots,k_r)} V^*$) 
the $(k_1,\dots,k_r)$-isotypical component of $V^{\ot n}$ (resp.\ $(V^*)^{\ot n}$),
i.e. the sum of all subrepresentations of $V^{\ot n}$ (resp.\ $(V^*)^{\ot n}$) 
isomorphic to $(k_1,\dots,k_r)$. We also put $\Lambda^n_R V = \Sigma^{(n)}_R V,$ 
$\Lambda^n_R V^* = \Sigma^{(n)}_R V^*$ for brevity.

\noindent{\sc Remark.}
The subspaces $\Sigma^\lambda_RV\subset V^{\ot n}$ are the $\Phi_R$-images of some objects 
of the category $\TT_R$.

Let $\lambda,$ $\mu$ be partitions of $n$ and $m$ respectively.
It is clear that the action of the permutation $\sigma_{n,m}\in\FS_{n+m}$
$$
\sigma_{n,m}(i)=\begin{cases}i+m, & \text{if $1\le i\le n$}\\i-n, & \text{if 
$n+1\le i\le n+m$}\end{cases}
$$
gives an isomorphism
$$
R_{n,m}:\Sigma^\lambda_RV\ot\Sigma^\mu_RV \to \Sigma^\mu_RV\ot\Sigma^\lambda_RV.
$$

\noindent{\sc Remark.}
This isomorphism is the image of an isomorphism in the category $\TT_R$.

The trivial example of a Yang-Baxter operator is the usual transposition
$$
R_0(v_1\ot v_2)=v_2\ot v_1.
$$
We will say that $R$ is a deformation-trivial Yang-Baxter operator if $R$ is
an algebraic deformation of $R_0$ in the class of Yang-Baxter operators.
For a deformation-trivial Yang-Baxter operator $R$ we have
$$
\dim\Sigma^\lambda_R V = \dim\Sigma^\lambda_{R_0} V
$$
for any partition $\lambda$.

\subsection{The noncommutative projective space}

Let $R$ be a 
deformation-trivial 
Yang-Baxter operator on the vector space $V^*$. Then the graded algebra
$$
S^\cdot_RV^* = T(V^*)\left/\Big\langle\Lambda^2_R V^*\Big\rangle\right.
$$
is a noncommutative deformation of the coordinate
algebra of the projective space $\PP(V)$. We denote
by $\PP_R(V)$ the corresponding noncommutative variety.
Thus $\PP_R(V)$ is a noncommutative
deformation of the projective space $\PP(V)$.

\begin{example}
The operator  
\begin{equation}\label{rab}
\begin{array}{l}\displaystyle
R(z_{i}\ot z_{j})= z_{j}\ot z_{i},\qquad\qquad\qquad\qquad 
\text{if $(i,j)\ne (1,2),$ $(2,1)$}\\\displaystyle
R(z_{1}\ot z_{2})=z_{2}\ot z_{1} + 2\h(az_3\ot z_4 + bz_4\ot z_3)\\
R(z_{2}\ot z_{1})=z_{1}\ot z_{2} - 2\h(bz_3\ot z_4 + az_4\ot z_3).
\end{array}
\end{equation}
is a deformation trivial Yang-Baxter operator on the $4$-dimensional
vector space $Z^*$ with the basis $\{z_1,z_2,z_3,z_4\}$.
By definition the homogeneous coordinate algebra of $\PP_R(Z)$ is generated 
by $z_1,z_2,z_3,z_4$ with relations $(\ref{rel})$ (we set $a+b=1$ as before).
Hence $\PP_R(Z)$ is isomorphic to the noncommutative projective space 
$\PT_\h$ defined in section~\ref{varieties}. The space $Z^*$ was denoted $U$ in that 
section.
\end{example}

The above example shows that part of the data encoded
in the Yang-Baxter operator~$R$ is lost in the structure
of the corresponding noncommutative projective space.
We will see below that this data appears
in the structure of other noncommutative varieties
associated with~$R$.

\subsection{Noncommutative Grassmannians}

It is well known that the homogeneous coordinate algebra of 
the Grassmann variety $\Gr(k;V)$ is a graded quadratic algebra
with $\Lambda^kV^*$ as the space of generators and
$$
\Ker\Big(\Lambda^kV^*\ot\Lambda^kV^*\to (V^*)^{\ot 2k} \to 
\Sigma^{(k,k)}V^*\Big)
$$
as the space of relations. This description justifies the following definition.

\begin{definition}
Let $R$ be a Yang-Baxter operator on the space $V^*$.
The noncommutative Grassmann variety $\Gr_R(k;V)$ is the noncommutative
projective variety corresponding to the quadratic algebra
$$
\gr_R(k;V) = T(\Lambda^k_R V^*)\left/
\Big\langle\Ker(\Lambda^k_R V^*\ot\Lambda^k_R V^*\to\Sigma^{(k,k)}_R 
V^*)\Big\rangle\right.
$$
\end{definition}
The algebra $\gr_R(k;V)$ is the $\Phi_R$-image of a commutative algebra
in the category $\TT_R$.

If $R$ is deformation-trivial, then $\Gr_R(k;V)$ is a noncommutative deformation of $\Gr(k;V)$.
Note that $\Gr_R(1;V)=\PP_R(V)$ by definition.

\begin{example}
Consider the noncommutative Grassmannian $\Gr_R(2;Z)$ corresponding to the 
Yang-Baxter operator $(\ref{rab})$. Let 
$$
z_{ij}=\frac12((z_i\ot z_j - z_j\ot z_i) - R(z_i\ot z_j - z_j\ot z_i)) \in 
\Lambda^2_R Z^*.
$$
Then it is easy to check that $\gr_R(2;Z)$ is generated by the elements 
$$
Y_1 = z_{13},\quad
Y_2 = -z_{24},\quad
Y_3 = z_{23},\quad
Y_4 = z_{14},\quad
D = -z_{12},\quad
T = z_{34},
$$
with relations
\begin{equation}\label{quadrelY}
\begin{array}{l}
\begin{array}{lcllcl}
{}[Y_1,Y_2] & = & \hphantom{-}2{\h}aT^2,  & [Y_3,Y_4] & = & 2{\h}bT^2, \\
{}[D,Y_1] & = & -2{\h}aY_1 T, & [D,Y_2] & = & 2{\h}aY_2 T, \\
{}[D,Y_3] & = & -2{\h}bY_3 T, & [D,Y_4] &=& 2{\h}bY_4 T, 
\end{array} \\
{}\qquad DT = \frac12 \left(Y_1Y_2+Y_2Y_1+Y_3Y_4+Y_4Y_3\right),
\end{array}
\end{equation}
$[Y_i,Y_j]=[T,Y_j]=[T,D]=0$ for all $i=3,4,\ j=1,2,3,4.$
Comparing with $(\ref{quadrel})$ one can see that the algebra $\gr_R(2;Z)$
is isomorphic to $\Q_{\h}$ with $G$ and $\theta$ given by
$$
G=\frac{1}{2}\begin{pmatrix}
0 & 1 & 0 & 0 \\
1 & 0 & 0 & 0 \\
0 & 0 & 0 & 1 \\
0 & 0 & 1 & 0 \\
\end{pmatrix},
\qquad
\theta=2{\h}\begin{pmatrix}
0 & a & 0 & 0 \\
-a & 0 & 0 & 0 \\
0 & 0 & 0 & b \\
0 & 0 & -b & 0 
\end{pmatrix}.
$$
Note that the variables $X_i,\ i=1,2,3,4,$ used in section~\ref{pthree} to describe
the quadric are related to $Y_i,\ i=1,2,3,4,$ by the following formulas:
\begin{equation}\label{xandy}
\begin{array}{ll}
Y_1=X_2+\sqrt{-1}\ X_1,\quad & Y_2=-X_2+\sqrt{-1}\ X_1,\\
Y_3=X_4+\sqrt{-1}\ X_3,\quad & Y_4=-X_4+\sqrt{-1}\ X_3.
\end{array}
\end{equation}
\end{example}


\subsection{Products of Grassmannians and flag varieties}

Let $R$ be a Yang-Baxter operator on the vector space $V^*$.
Consider a sequence $k_1,\dots,k_r$ of integers.
Let $\ZZ^r$ be the free abelian group with $r$
generators $e_1,\dots,e_r$. The $R$-tensor product
$$
\gr_R(k_1;V)\otR\dots\otR\gr_R(k_r;V)
$$
is a $\ZZ^r$-graded algebra generated by the vector spaces $\Lambda^{k_i}_RV^*$
in degree $e_i$, with relations 
$$
\Ker\Big(\Lambda^{k_i}_R V^*\ot\Lambda^{k_i}_R V^*\to\Sigma^{(k_i,k_i)}_R 
V^*\Big)
$$
in degree $2e_i$ for all $i$ and
$$
\begin{CD}
\Ker\Big((\Lambda^{k_i}_R V^*\ot\Lambda^{k_j}_R V^*)\oplus(\Lambda^{k_j}_R 
V^*\ot\Lambda^{k_i}_R V^*)
@>{\quad(\id, - R_{k_j,k_i})\quad}>>
\Lambda^{k_i}_R V^*\ot\Lambda^{k_j}_R V^*\Big)
\end{CD}
$$
in degree $e_i+e_j$ for all $i>j$. For any increasing sequence $k_1,\dots,k_r$
we define also a $\ZZ^r$-graded algebra $\fl_R(k_1,\dots,k_r;V).$ It has the
same generators as the algebra $\gr_R(k_1;V)\otR\dots\otR\gr_R(k_r;V),$,
subject to the same relations in degrees $2e_i$ and to relations 
$$
\begin{CD}
\Ker\Big((\Lambda^{k_i}_R V^*\ot\Lambda^{k_j}_R V^*)\oplus(\Lambda^{k_j}_R 
V^*\ot\Lambda^{k_i}_R V^*)
@>(\id ,- R_{k_j,k_i})>>
\Lambda^{k_i}_R V^*\ot\Lambda^{k_j}_R V^*
@>>>\Sigma^{(k_i,k_j)}_R V^*\Big)
\end{CD}
$$
in degree $e_i+e_j$ for all $i>j$. 
This definition is suggested by the Borel-Weil-Bott theorem
(see~\cite{BBW}). In particular, for $R=R_0$ we get the algebra
corresponding to the commutative flag variety.


We define the $R$-Carthesian product
$\Gr_R(k_1;V)\timesR\dots\timesR\Gr_R(k_r;V)$ and the noncommutative flag 
variety
$\Fl_R(k_1,\dots,k_r;V)$ as the noncommutative varieties corresponding
to the algebras $\gr_R(k_1;V)\otR\dots\otR\gr_R(k_r;V)$ and 
$\fl_R(k_1,\dots,k_r;V)$ respectively. 

To make this compatible with our definition of a noncommutative variety,
we consider instead of a $\Z^r$-graded algebra its diagonal subalgebra.
The diagonal subalgebra is a graded algebra whose $n$-th graded component
is the $n(e_1+\dots+e_r)$-graded component of the $\Z^r$-graded algebra.
Thus according to section~\ref{varieties} the category of
coherent sheaves on the $R$-Cartesian product of Grassmannians (or the flag variety) 
is the category $qgr$ of the corresponding diagonal subalgebra.

The algebra $\fl_R(k_1,\dots,k_r;V)$ is the $\Phi_R$-image of a commutative algebra
in the category $\TT_R$. 
Hence one can define the $R$-Carthesian product of several flag varieties. 

If $R$ is deformation-trivial, then 
$\Gr_R(k_1;V)\timesR\dots\timesR\Gr_R(k_r;V)$ and $\Fl_R(k_1,\dots,k_r;V)$ 
are noncommutative deformations of the corresponding commutative varieties.

Note that we have a canonical embedding of the graded algebra 
$\gr_R(k_i;V)$ into the graded algebra $\fl_R(k_1,\dots,k_i,\dots,k_r;V)$ inducing 
the canonical projections 
$$
p_i:\Fl_R(k_1,\dots,k_i,\dots,k_r;V)\to\Gr_R(k_i;V).
$$
On the other hand, by definition $\fl_R(k_1,\dots,k_r;V)$ is a quotient algebra
of the algebra $\gr_R(k_1;V)\otR\dots\otR\gr_R(k_r;V)$.
Hence $\Fl_R(k_1,\dots,k_r;V)$ can be regarded as a closed subvariety in
$\Gr_R(k_1;V)\timesR\dots\timesR\Gr_R(k_r;V)$.

\begin{example}\label{fl12}
The algebra $\gr_R(1;Z)\otR\gr_R(2;Z)$ corresponding to the Yang-Baxter 
operator $(\ref{rab})$
is generated by the elements $z_1,z_2,z_3,z_4,Y_1,Y_2,Y_3,Y_4,D,T$ with 
relations $(\ref{rel})$, $(\ref{quadrelY}),$ and
$$
\begin{array}{lcllcl}
{}[z_1,Y_2] &=& - 2{\h}az_4T,            & [z_2,Y_1] &=& \hphantom{-}2{\h}az_3T, \\
{}[z_1,Y_3] &=& - 2{\h}bz_3T,            & [z_2,Y_4] &=&           - 2{\h}bz_4T, \\
{}[z_1,D]   &=& - 2{\h}bz_3Y_4-2{\h}az_4Y_1, & [z_2,D] &=& \hphantom{-}2{\h}az_3Y_2-2{\h}bz_4Y_3,
\end{array}
$$
$[z_1,Y_1]=[z_2,Y_2]=0,$ $[z_3,Y_i]=[z_3,D]=0,$ $[z_4,Y_i]=[z_4,D]=0,$ $[z_i,T]=0$ for all 
$i=1,2,3,4$. 
The algebra $\fl_R(1,2;Z)$ is given by the same generators subject to the same
relations, as well as the additional relations
\begin{equation}\label{fl12eq}
\arraycolsep=2pt
\begin{pmatrix}
     0  &  T  &  Y_2 &               Y_3 \\
T   &       0  & -  Y_4 &               Y_1 \\
Y_2 & Y_4 &       0  &        D-{\h}(a+b)T \\
Y_3 & - Y_1 & -D-{\h}(a+b)T &                0  \\
\end{pmatrix}
\begin{pmatrix}
z_1\\
z_2\\
z_3\\
z_4\\
\end{pmatrix}
=
\begin{pmatrix}
0\\
0\\
0\\
0\\
\end{pmatrix}.
\end{equation} 
As explained above, we have projections
$$
\begin{CD}
\Q_\h @= \Gr_R(2;Z) @<{p}<< \Fl_R(1,2;Z) @>{q}>> \PP_R(Z) @= 
\PT_{\h}
\end{CD}
$$
and a closed embedding
$$
\Fl_R(1,2;Z) \subset \Gr_R(2;Z)\timesR \PP_R(Z) = 
\Q_\h\timesR\PT_{\h}.
$$
\end{example}

\subsection{Tautological bundles}

Let $\CV$ (resp.\ $\CV^*,$ $\Sigma^\lambda_R\CV,$ $\Sigma^\lambda_R\CV^*$) 
denote the coherent sheaf 
on $\Gr_R(k;V)$ corresponding to the free right $\gr_R(k;V)$-module 
$V\ot\gr_R(k;V)$ 
(resp.\ $V^*\ot\gr_R(k;V),$ $\Sigma^\lambda_RV\ot\gr_R(k;V),$ 
$\Sigma^\lambda_RV^*\ot\gr_R(k;V)$).
Since the space of global sections of the sheaf $\CO(1)$ on
the Grassmannian $\Gr_R(k;V)$ is $\Lambda^k_RV^*$,
the maps $\Lambda^{k-1}_RV^*\to V\ot\Lambda^k_RV^*$ and $\Lambda^{k+1}_RV^*\to 
V^*\ot\Lambda^k_RV^*$ 
induce morphisms of sheaves
$$
\begin{CD} 
\Lambda^{k-1}_R\CV^*(-1) @>{\phi}>> \CV    \qquad \text{and} \qquad
\Lambda^{k+1}_R\CV^*(-1) @>{\psi}>> \CV^*.
\end{CD}
$$
We put $S=\Image\phi,$ $\CV/S=\Coker\phi$, 
$\Spr = \Image\psi,$ 
$\CV^*/\Spr=\Coker\psi$.

\medskip
\noindent{\sc Remark.}
For $k=1$ we have $S=\CO(-1),$ $\CV^*/\Spr=\CO(1)$.
\medskip

One can show that these sheaves are locally free.
We refer to them as tautological bundles. 

The free $\gr_R(k;V)$-modules, corresponding to the sheaves $\Sigma^\lambda_R\CV$,
$\Sigma^\lambda_R\CV^*$ are the $\Phi_R$-images of free modules over the corresponding
algebra in the category $\TT_R$. Furthermore, the morphisms $\phi$ and $\psi$
are $\Phi_R$-images. This implies that the $\gr_R(k;V)$-modules corresponding
to the tautological bundles are $\Phi_R$-images as well. Therefore they all 
have a natural structure of $\gr_R(k;V)$-bimodules. This allows to define
$R$-symmetric powers $S^k_R(-)$ (resp.\ $R$-exterior powers $\Lambda^k_R(-)$) 
of the tautological bundles as the corresponding $\Phi_R$-images.

One can check that we have canonical isomorphisms of bimodules
$$
\CV^*/\Spr \cong S^\vee,\qquad \Spr \cong (\CV/S)^\vee.
$$

\begin{example}\label{phipsi}
Let $R$ be the Yang-Baxter operator $(\ref{rab})$ and $k=2$.
Let $\cz_1,\cz_2,\cz_3,\cz_4$ be the dual basis of $Z$.
Then the twisted maps
\begin{eqnarray}
&&\phi(1):Z^*\ot \CO_{\Gr_R}\to Z\ot \CO_{\Gr_R}(1),\nonumber\\
&&\psi(1):Z\ot \CO_{\Gr_R}\cong\Lambda^3_RZ^*\ot 
\CO_{\Gr_R}\to Z^*\ot\CO_{\Gr_R}(1)\nonumber
\end{eqnarray}
are given by
$$
\begin{array}{cccc}
\phi(1):
\begin{pmatrix}
z_1\\
z_2\\
z_3\\
z_4\\
\end{pmatrix}
&
\mapsto
&
\begin{pmatrix}
   0  & D+{\h}(a-b)T &  -Y_1 & -Y_4 \\
  D-{\h}(a-b)T & 0 & -Y_3 &  Y_2\\
-Y_1 & Y_3 & 0 & -T   \\
-Y_4 & - Y_2 & T & 0   \\ 
\end{pmatrix}
&
\begin{pmatrix}
\cz_1\\
\cz_2\\
\cz_3\\
\cz_4\\
\end{pmatrix}
\\
\\
\psi(1):
\begin{pmatrix}
\cz_1\\
\cz_2\\
\cz_3\\
\cz_4\\
\end{pmatrix}
&
\mapsto
&
\begin{pmatrix}
0 & T    & Y_2 &              Y_3 \\
 T  & 0 & -Y_4 &              Y_1 \\
Y_2 & Y_4 & 0 & D-{\h}(a+b)T \\
Y_3 & -Y_1 & -D-{\h}(a+b)T & 0\\ 
\end{pmatrix} 
&
\begin{pmatrix}
z_1\\
z_2\\
z_3\\
z_4\\
\end{pmatrix}
\end{array}
$$
Note that $\psi(1)\phi=0$ and $\phi(1)\psi=0$. Hence we have isomorphisms
$$
\Spr(1)\cong \CV/S,\qquad S(1)\cong \Ssta.
$$
Note also that on the open subset $T\ne0$ elements $(z_3,z_4)$ 
give a trivialization of the tautological bundle $\Ssta$. 
More precisely, the restriction of the sections $z_1,z_2$ of $\Ssta$
can be expressed as
\begin{equation}\label{triv}
z_1 = y_4z_3-y_1z_4,\qquad z_2=-y_2z_3-y_3z_4
\end{equation}
where $y_i=T^{-1}Y_i$. Similarly, the elements$(\cz_1,\cz_2)$
give a trivialization of $\CV/S$ on $T\ne0$. Thus the restrictions 
of all tautological bundles to the open subset $T\ne0$ correspond 
to the free rank two bimodule over the Weyl algebra $A(\C^4_{\h}).$
\end{example}

\subsection{Pull-back and push-forward}

Recall that we have canonical projections 
$p_i:\Fl_R(k_1,k_2;V)\to\Gr_R(k_i;V)$ ($i=1,2$).
Given a right graded $\gr_R(k_i;V)$-module $E$ we consider
the right bigraded $\fl_R(k_1,k_2;V)$-module
$E\ot_{\gr_R(k_i;V)}\fl_R(k_1,k_2;V)$.
The diagonal subspace of this module is a graded 
module over the diagonal subalgebra of $\fl_R(k_1,k_2;V)$.
This gives the pull-back functor
$$
p_i^*:coh(\Gr_R(k_i;V))\to coh(\Fl_R(k_1,k_2;V)). 
$$
The pull-back functor is exact and takes a $\Phi_R$-image to a $\Phi_R$-image.
In particular, the pull-backs of the tautological bundles
have a canonical bimodule structure.

The pull-back functor $p_i^*$ admits a right adjoint functor 
$p_{i*}:coh(\Fl_R(k_1,k_2;V))\to coh(\Gr_R(k_i;V)),$ called the 
push-forward functor. It also takes a $\Phi_R$-image to a $\Phi_R$-image.


The line bundles $p_1^*\CO(i)$ and $p_2^*\CO(j)$ on the flag variety $\Fl_R(k_1,k_2;V)$
are $\Phi_R$-images, hence they have a canonical bimodule structure. Therefore,
we have a well-defined tensor product
$$
\CO(i,j)=p_1^*\CO(i)\ot p_2^*\CO(j).
$$
The line bundle $\CO(i,j)$ is also a $\Phi_R$-image and has a canonical bimodule structure. 
The $n$-th graded component
of the corresponding module over the diagonal subalgebra of $\fl_R(k_1,k_2;V)$ is 
the $((n+i)e_1+(n+j)e_2)$-graded component of the algebra $\fl_R(k_1,k_2;V)$.


One can check that the push-forward of the line bundle $\CO(j_1,j_2)$
with respect to $p_2$ is given by the formula
$$
p_{2*}\CO(j_1,j_2)=S^{j_1}_R(\Ssta)(j_2).
$$


\subsection{$\Fl_R(1,2;Z)$ as the projectivization of the tautological bundle}

The $R$\!--symmetric powers of the tautological bundle form a sheaf of graded algebras 
on the Grassmannian $\Gr_R(k;V)$ 
$$
S^\cdot_R(\Ssta) = T(\Ssta)\left/\Big\langle\Lambda^2_R\Ssta\Big\rangle\right..
$$
The corresponding $\gr_R(k;V)$-module 
$$
\bigoplus_{i,j=0}^\infty\Gamma(\Gr_R(k;V),S^j_R(\Ssta)(i))
$$
is a bigraded module with a structure of a bigraded algebra.
One can check that this bigraded algebra is isomorphic 
to the bigraded algebra $\fl_R(1,k;V)$. Thus we can regard
the flag variety $\Fl_R(1,k;V)$ as the projectivization of the
tautological bundle $S$ on the Grassmannian $\Gr_R(k;V)$.
In particular, $\Fl_R(1,2;Z)$ is the projectivization of the
tautological bundle $S$ on the Grassmannian $\Gr_R(2;Z)$.


\subsection{Noncommutative twistor transform}

If $E$ is a coherent sheaf on the noncommutative projective space 
$\PP_R(Z)=\PTh,$ we define its twistor transform as the sheaf $p_*q^*E$ on 
$\Gr_R(2;Z)=\Q_\h$, where $q$ is the projection $\Fl_R(1,2;Z)\to\PP_R(Z)=\PTh$
and $p$ is the projection $\Fl_R(1,2;Z)\to\Gr_R(2;Z)=\Q_\h$. Similarly, 
we can define the twistor transform of a complex of sheaves on $\PTh$. 
Actually, it is more
natural to consider the derived twistor transform, i.e.\ the derived
functor of the ordinary twistor transform.

Consider a complex $\Ko$ of the form
$$
0\lto H\ot \O(-1)\stackrel{M}{\lto} K\ot \O \stackrel{N}{\lto} L\ot \O(1)\lto 0
$$
on the projective space $\PTh$. One can check that under the twistor transform one has
$$
\CO_{\PTh}(-1)\mapsto 0,\qquad\CO_{\PTh}\mapsto\CO_{\Gr_R},\qquad\CO_{\PTh}(1)\mapsto\Ssta.
$$
In fact, for these sheaves the derived twistor transform coincides with the ordinary one.
Thus the (derived) twistor transform takes the complex $\Ko$
to the complex
$$
0 \lto K\ot\CO \stackrel{\New}{\lto} L\ot\Ssta \lto 0.
$$

Let $\E$ denote the middle cohomology of the complex $\Ko$.
It follows that the twistor transform takes $\E$ to 
the kernel of the map $\New: K\ot\CO \lto L\ot\Ssta$.

One can describe $\New$ without reference to the twistor transform. 
The morphism $N$ is the same thing as a vector space morphism
\begin{equation}\label{Ntensor}
N_1z_1+N_2z_2+N_3z_3+N_4z_4:\ K\lto Z^*\ot L.
\end{equation}
Here the maps $N_i$ are given in terms of the deformed ADHM data according 
to~(\ref{form1}) and (\ref{form2}).
The map $\New$ is a composition of two maps
$$
K\ot \CO_{\Gr_R}\lto L\ot Z^*\ot \CO_{\Gr_R}\lto L\ot S^{\vee},
$$
where the first map is given by~(\ref{Ntensor}), while the second map comes from
the canonical projection $Z^*\ot \CO_{\Gr_R}\to S^{\vee}$. (We remind that 
$S^{\vee}$ is the cokernel of the map $\psi: Z\ot \CO_{\Gr_R}(-1)\lto Z^*\ot \CO_{\Gr_R}.$)

Recall that on the open subset $\{T\ne0\}$ the bundle $\Ssta$ is trivial, and 
the elements $(z_3,z_4)$ give its trivialization (see (\ref{triv})). Hence the restriction 
of the twistor transform of the complex~(\ref{comp}) to this open subset 
is isomorphic to the complex
\begin{equation}\label{comp2}
\begin{CD}
0 @>>> K\ot\CO @>{\begin{pmatrix} N_3+y_4N_1-y_2N_2\\
N_4-y_1N_1-y_3N_2\end{pmatrix}}>>
(L\oplus L)\ot\CO \lto 0.
\end{CD}
\end{equation}

Assume now that the complex~(\ref{comp}) is given by the deformed ADHM data
$(\Bnul,\Bone,\Nu,\Mu)$ (see section~(\ref{pthree})).
Applying the formulas~(\ref{form1}) and (\ref{form2}), we see that with respect
to the chosen bases of $L$ and $K$ the map $\New$ is given by the matrix
$$
\begin{pmatrix}
-\Bone+y_2 & \Bnul+y_4 & \Nu \\
-\Bnul^\dag+y_3 & -\Bone^\dag-y_1 & -\Mu^\dag
\end{pmatrix}.
$$
It is evident that this operator is related to the operator ${\mathcal D}$ in (\ref{D})
by a change of basis. In particular, the Nekrasov-Schwarz coordinates $\xi_1,\xi_2,
\bxi_1,\bxi_2$ (see section~\ref{summary}) can be expressed through $x_i=T^{-1}X_i$ as 
follows:
$$
\begin{array}{ll}
\xi_1=-y_4=x_4-\sqrt{-1}\ x_3, & \xi_2=y_2=-x_2+\sqrt{-1}\ x_1,\\
\bxi_1=y_3=x_4+\sqrt{-1}\ x_3, & \bxi_2=-y_1=-x_2-\sqrt{-1}\ x_1.
\end{array}
$$
Thus the twistor transform of the complex corresponding to the deformed ADHM data
coincides with the instanton bundle corresponding to these data (see section~\ref{summary}).
This gives a geometric interpretation of the deformed ADHM construction of the noncommutative
instanton bundle.

\subsection{Differential forms}\label{differentials}

Let an algebra $A$ be the $\Phi_{R}$\!--image of a commutative algebra
in the category $\TT_{R}$. 
This means that there exists an operator $R: A^{\ot 2}\lto A^{\ot 2}$
compatible with the multiplication law of $A$.
Above we have defined the $R$\!--tensor product $A\otR A$ which is also
an algebra with a Yang-Baxter operator. Explicitly, the multiplication law of
$A\otR A$ is defined as follows. Let $m$ be the multiplication map from $A\ot A$
to $A$. Then the multiplication map from $(A\ot A)\ot (A\ot A)$ to $A\ot A$
is given by 
$m_{12} m_{34} R_{23}$
in the obvious notation.
It is easy to see that the multiplication map $m$ is a homomorphism of algebras.

Let $I$ denote the kernel of the map $m:A\otR A\to A$.
Then $I$ is a two-sided ideal of the algebra $A\otR A$.

\begin{definition}
We define the bimodule of $R$\!--differential forms of the algebra $A$ by
$$
\Om^1_A =I/I^2 .
$$
\end{definition}
\noindent For a motivation of this definition, see~\cite{Connes}.
Furthermore, suppose $A$ is a graded algebra. 
Consider the total grading of the bigraded algebra $A\otR A$.
The two-sided ideal $I$ inherits the grading. Therefore the bimodule $\Om_{A}^1$ is graded
too.

In the graded case, besides $\Om^1_{A},$ 
we can define the module of projective differential forms of $A$ in the following way.
Let $\chi: A\otR A\to A
\otR A$ be the linear 
operator which acts on the $(p,q)$\!--th graded component of 
the algebra $ A\otR A$ as a scalar multiplication by $q$.
Since $\chi$ is a derivation, we have $\chi(I^2)\subset I$. Therefore 
$m(\chi(I^2))=0$.
Furthermore, the induced map 
$\Om^1_{A}=I/I^2\stackrel{m\cdot\chi}{\lto} A$ 
is a morphism of graded $A$\!--bimodules.
\begin{definition}
We define the $A$\!--bimodule of projective differential forms of the algebra $A$ by
$$
\wh{\Om}^1_{A} = 
\Ker(\Om^1_{A}\stackrel{m\cdot\chi}{\lto} A).
$$
\end{definition}

First, let us apply this construction of differential forms to the noncommutative
affine variety $\C^4_{\h}$ (subsection \ref{noncomC}). 
The algebra $A(\C^4_{\h})$ of polynomial functions on
$\C^4_{\h}$ is the Weyl algebra:
$$
A(\C_{\h}^4) = \T(x_1,x_2,x_3,x_4)/\langle[x_i,x_j]=
\h\theta_{ij}\rangle _{1\le i,j \le 4}.
$$
Let us define the Yang-Baxter operator on the tensor square of the subspace of $A(\C_{\h}^4)$
spanned by $1,x_1,x_2,x_3,x_4$ by the formula
$$
1\ot x_i\mapsto x_i\ot 1,\quad 
x_i\ot 1 \mapsto 1\ot x_i,\quad
x_i \ot x_j \mapsto x_j \ot x_i + \h\theta_{ij}\cdot 1\ot 1 \quad
\text{for all}\quad
1\le i, j \le 4.
$$
This Yang-Baxter operator has a unique extension to the whole $A(\C_{\h}^4)$ compatible
with the multiplication law. 

There is another way to look at this Yang-Baxter operator. Recall that $\C_{\h}^4$ is
an open subset $T\neq 0$ in the noncommutative Grassmannian $\Gr_R(2;Z)$ where $R$
is defined by (\ref{rab}).
The Yang-Baxter operator on the quadratic algebra $\gr_R(2;Z)$ has the property
that $R(T\ot a)=a\ot T$ for any $a\in \gr_R(2;Z)$. Hence it descends to a Yang-Baxter
operator on $A(\C_{\h}^4)$. It is easy to see that it acts on the tensor square of the
subspace spanned by $1,x_1,x_2,x_3,x_4$ in the above manner.

We define the sheaf of differential forms $\Omega^1_{\C^4_{\h}}$ as the bimodule of 
$R$\!--differential forms of the algebra $A(\C^4_{\h})$. 
It is easy to check that $\Omega^1_{\C^4_{\h}}$ is isomorphic to the bimodule
$A(\C^4_{\h})^{\oplus 4}$.
Futhermore, we can take any $R$\!--exterior power of $\Omega^1_{\C^4_{\h}}$ and
thereby define $\Omega^p_{\C^4_{\h}}$.
This enables us to define a connection and  its curvature
on any bundle on the noncommutative
affine space. The relevant formulas were written above (see subsection \ref{noncomminst}).

Second, we define the sheaf of differential forms $\Omega^1_{\Gr_R}$ on the noncommutative 
Grassmannian $\Gr_R(k;V)$
as the sheaf corresponding to the module of projective differential forms 
$\wh{\Om}^1_{\gr_R}$.

It can be shown that as in the commutative case we have an isomorphism of coherent
sheaves on the noncommutative Grassmannian $\Gr_R(k;V)$:
$$
\Omega^1_{\Gr_R}\cong S\ot \Spr.
$$
It follows that for $k=1$  that we have an exact sequence 
$$
0 \longrightarrow \Omega^1_{\PP_R(V)} \longrightarrow \CV^*(-1) \longrightarrow 
\CO \longrightarrow 0.
$$
Thus this definition of the sheaf of differential forms $\Omega^1_{\PP_R(V)}$
is consistent with Definition~\ref{Omk}.

Similarly, 
one can define the sheaf of differential forms $\Omega^1_{\Fl_R}$ on the noncommutative 
flag variety 
$\Fl_R(k_1,\dots,k_r;V)$. One can check that the projection
$$
p_i:\Fl_R(k_1,\dots,k_i,\dots,k_r;V)\to\Gr_R(k_i;V)
$$
induces a morphism of
bundles  
$p_i^*: \Omega^1_{\Gr_R}\to\Omega^1_{\Fl_R}$.

In the commutative case the ADHM construction of the instanton connection can be interpreted
in terms of twistor transform (see \cite{Atiyah, Manin} for details).
We believe
that this can be done in the noncommutative case as well.
It appears that the most convenient definition of connection on a bundle
on a noncommutative projective variety
is in terms of jet bundles (see, for example, \cite{Manin}).

\section{Instantons on a $q$\!--\,deformed $\RR^4$}\label{sec:qdeformed}

In this paper we have focused on a particular noncommutative deformation of $\RR^4$ 
related to the Wigner-Moyal product~(\ref{wm}). This is the only deformation
of $\RR^4$ which is known to arise in string theory. But most of our constructions
work for more general deformations which do not have a clear physical
interpretation. For example, let us replace
$\CC^4_{\h}$ with a noncommutative affine variety whose coordinate ring
is generated by $z_1,z_2,z_3,z_4$ subject to the following quadratic relations:
$$
q z_1 z_2-q^{-1} z_2 z_1=\h,\quad q z_3 z_4 -q^{-1} z_4 z_3=\h,\quad [z_1,z_3]=[z_1,z_4]=
[z_2,z_3]=[z_2,z_4]=0.
$$
We will denote this noncommutative affine variety by $\CC^4_{q,\h},$
and its coordinate algebra by ${\mathcal A}_{q,\h}$.
If $\h$ and $q$ are real, we can define a $*$-operation on  ${\mathcal A}_{q,\h}$
by $z_1^*=z_2,\ z_3^*=z_4$. The corresponding real noncommutative affine variety will 
be denoted by $\RR^4_{q,\h}.$

Consider now the following deformation of the ADHM equations:
\begin{equation}
[B_1,B_2]_{q^{-1}}+IJ  =  0,\quad
[B_1,B_1^\dag]_{q^{-1}}+[B_2,B_2^\dag]_q+II^\dag-J^\dag J  =  -2\h\cdot 1_{k\times k}.
\end{equation}
Here $B_1,B_2\in \Hom(V,V),$ $I\in \Hom(W,V),$ $J\in \Hom(V,W)$, as usual, and
by $[A,B]_q$ we mean a $q$-commutator:
$$
[A,B]_q=qAB-q^{-1}BA.
$$
We claim that solutions of these ``$q$\!--\,deformed'' ADHM equations can be used as an 
input for the construction of instantons on $\RR^4_{q,\h}$ of rank $r=\dim W$ and instanton 
charge $k=\dim V$. Let us sketch this construction. Define an operator 
$${\mathcal D}\in \Hom_{{\mathcal A}_{q,\h}} ((V\oplus V\oplus W)\ot_\CC{{\mathcal A}_{q,\h}},
(V\oplus V)\ot_\CC{{\mathcal A}_{q,\h}})$$
by the formula
$$
{\mathcal D}=\begin{pmatrix}
B_1-q z_1 & -q B_2+q z_2 & I \\
B_2^\dag-\bz_2 & q B_1^\dag-\bz_1 & J^\dag 
\end{pmatrix}.
$$
Now we can go through the same manipulations as in section~\ref{summary}: 
assume that ${\mathcal D}$ is surjective, and its kernel is a 
free module, and define a connection 1-form by the expression~(\ref{asdconnection}). 
The same formal computation as in section~\ref{summary} shows that the curvature
of this connection is anti-self-dual. 

In order to ensure that ${\mathcal D}$ is surjective, 
it is probably necessary to replace the algebra ${{\mathcal A}_{q,\h}}$ with some bigger 
algebra containing ${{\mathcal A}_{q,\h}}$ as a subalgebra. This bigger algebra should play 
the role of the algebra of smooth functions on our noncommutative $\RR^4$. For $\h=0$,
$q\neq 1$ there is even a natural candidate for this bigger algebra: it should consist
of $C^\infty$ functions on $\CC^2$ with some suitable growth conditions at infinity
and the product defined by
\begin{multline}
(f\star g)(z_1,z_2,\bz_1,\bz_2)=\\
\exp\left(-\ln(q)\left(z_1 \bz_1'\frac{\partial^2}{\partial z_1\partial\bz_1'}
+z_2 \bz_2'\frac{\partial^2}{\partial z_2\partial\bz_2'}-
z_1'\bz_1\frac{\partial^2}{\partial z_1'\partial\bz_1}-
z_2'\bz_2\frac{\partial^2}{\partial z_2'\partial\bz_2}
\right)\right)\\ f\left(z_1,z_2,\bz_1,\bz_2\right) g\left(z_1',z_2',\bz_1',\bz_2'\right)
\vert_{z_1'=z_1,z_2'=z_2}.
\end{multline}
Assuming that this formal expression exists, it is easy to check that the product is 
associative, that polynomial functions form a subalgebra with respect to it,
and that this subalgebra is isomorphic to ${{\mathcal A}_{q,\h}}$. 

It is natural to conjecture that all instantons
on $\RR^4_{q,\h}$ arise from this deformed ADHM construction. Note that in this
case the deformed ADHM equations are not hyperk\"ahler moment map equations, and
one cannot use the hyperk\"ahler quotient construction to infer the existence of
hyperk\"ahler metric on the quotient space.

The algebro-geometric part of the story can also be generalized. We did not go through
this carefully, but nevertheless would like to indicate one result. It appears that
the $q$\!--\,deformed ADHM data can be interpreted in terms of sheaves on a 
more general noncommutative $\PP^2$ than the one defined in section~\ref{varieties}. 
The graded algebra corresponding to this noncommutative $\PP^2$ is generated by degree 
one elements $z_1,z_2,z_3$ with the quadratic relations
$$
q z_1 z_2 - q^{-1} z_2 z_1= 2\h z_3^2, \quad [z_i,z_3]=0,\ i=1,2.
$$
This algebra is one of the Artin-Schelter regular algebras of dimension 
three~\cite{AS,ATV}.
It is characterized by the fact that the corresponding noncommutative variety 
$\PP^2_{q,\h}$ contains as subvarieties a commutative quadric and a noncommutative line.
The latter is given by the equation $z_3=0$. In the limit $q\ra 1$ the plane 
$\PP^2_{q,\h}$ reduces to $\PP^2_\h$, and the union 
of the quadric and the line turns into the triple commutative line $l$ which played such a prominent
role in this paper. If $q\neq 1$, then in the limit $\h\ra 0$ the quadric turns into a union 
of two intersecting commutative lines $z_1=0$ and $z_2=0$. 

For any $q$ the line $z_3=0$ should be regarded as 
``the line at infinity'' (which is noncommutative for $q\neq 1$). It is plausible that 
the $q$\!--\,deformed ADHM data are in one-to-one correspondence with bundles, 
or maybe torsion--free sheaves, on $\PP^2_{q,\h}$ with a trivialization on this line.

\section{Appendix}
In this section we define a $\star$--product on the space of complex-valued 
$C^\infty$ functions on $\RR^n$
whose derivatives of arbitrary order are polynomially bounded. 
The $\star$--product endows this space with a structure of a $\CC$-algebra and reduces to 
the Wigner-Moyal product~(\ref{wm}) on polynomial functions.

\begin{definition} Let $\Phi$ be a topological vector space which is a subspace of the space of
$C^\infty$ functions on $\RR^n,$ and let $\Phi'$
be the space of distributions on $\Phi$. Let $f$ be a $\CC$-valued function
on $\RR^n$ which simultaneously is a distribution in $\Phi'$. $f$ is called
a multiplier if for any $\phi\in \Phi$ $f\phi \in \Phi$.
\end{definition}
The set of multipliers of $\Phi'$ is obviously a subspace of $\Phi'.$

\begin{definition} Let $f\in\Phi'$.
$f$ is called a convolute if for any $\phi\in\Phi$ we have
$$
(f*\phi)(x)\equiv (f(\xi),\phi(x+\xi))\in \Phi,
$$
and this expression depends continuously on $\phi$. The above expression is
called the convolution of $f$ with $\phi$.
\end{definition}

The set of convolutes is obviously a subspace of $\Phi'.$

We will denote the Fourier duals of $\Phi$ and $\Phi'$ by $\widetilde{\Phi}$
and $\widetilde{\Phi'},$ respectively. If $f\in \Phi,$ then ${\widetilde f}\in
{\widetilde \Phi}$ will be the Fourier transform of $f,$ etc.

\begin{definition} The Schwartz space $\CS(\RR^n)$ is the space of $\CC$-valued 
$C^\infty$ functions on $\RR^n$ such that $\phi\in \CS$ if and only if all the norms
\be\label{norms}
\sup_x\ x^k D^m \phi(x),\qquad k=0,1,2,\ldots,
\end{equation}
are finite. Here $m=(m_1,\ldots,m_n)$ is an arbitrary polyindex.
\end{definition}

Convergence on $\CS$ is defined using the family of norms~(\ref{norms}). Then $\CS$
becomes a complete countably normed space~\cite{GS}.

\begin{proposition} A function $f\in \CS'$ is a multiplier if and only if it is a $C^\infty$
function on $\RR^n$ all of whose derivatives are polynomially bounded.
\end{proposition}
\begin{proof} Obvious.
\end{proof}

The following theorem proved in~\cite{Swartz} describes the subspace of convolutes of $\CS'$:
\begin{theorem}\label{structconv} A distribution $f\in \CS'$ is a convolute if and only if
it has the form
$$
f=\sum_{|\alpha|<r} D^\alpha f_\alpha(x),
$$
where $r$ is a positive integer, and  $f_\alpha$ are $C^0$ functions on $\RR^n$ which decrease 
at infinity faster than any negative power of $x$.
\end{theorem}
The functions which decrease at infinity faster than any negative power will be called
rapidly decreasing.

The following theorem is proved in~\cite{GS}, vol. 2, ch. III:
\begin{theorem} Fourier transform and its inverse act as automorphisms 
on both $\CS$ and $\CS'$.
\end{theorem}
{}From now on we identify $\CS\cong \widetilde{\CS},$ $\CS'\cong \widetilde{\CS'}.$

\begin{theorem}\label{isoconv} Fourier transform and its inverse establish an isomorphism
between the space of multipliers and the space of convolutes of $\CS'$.
\end{theorem} 
\begin{proof} By the preceding theorem, it is sufficient to show that the Fourier transform
of every multiplier is a convolute, and vice versa. The former fact is proved in \cite{GS}, 
vol.~2, ch.~III. Let us prove the converse.

By theorem~\ref{structconv}, every convolute has the form
$$
f(x)=\sum_{|\alpha|<r} D^\alpha f_\alpha(x)
$$
for some $r$ and rapidly decreasing continuous functions $f_\alpha$.
Let 
$$
{\widetilde f}_\alpha(p)=\int\ f_\alpha(x)\ e^{\sqrt{-1}px}\ d^n x.
$$
be the Fourier transform of $f_\alpha(x).$
Since the integrals
$$
\int\ x^\beta f_\alpha(x)\ e^{\sqrt{-1}px}\ d^n x
$$
are absolutely convergent, the functions ${\widetilde f}_\alpha$ are $C^\infty$ functions. 
Furthermore, the Fourier transform
of $f$ is equal to
$$
{\widetilde f}(p)=\sum_{|\alpha|<r} (-\sqrt{-1}\ p)^\alpha\ {\widetilde f}_\alpha(p)
$$
(see \cite{GS}, vol. 2, ch. III), hence ${\widetilde f}$ is also
a $C^\infty$ function. Finally, since by the preceding theorem the Fourier transform 
of any element of $\CS'$ is again an element of $\CS',$ ${\widetilde f}$ and all 
its derivatives are polynomially bounded. Hence ${\widetilde f}$ is a multiplier. 
\end{proof}

\begin{definition} Let $\omega$ be a skew-symmetric real-valued bilinear form on $\RR^n$.
The $\dia$--product on the space of convolutes of $\CS'$ is defined by
$$
({\widetilde f}\dia {\widetilde g})(p)=\int\ {\widetilde f}(q)\ {\widetilde g}(p-q)
\  e^{\sqrt{-1}\omega(p,q)}\ \frac{d^nq}{(2\pi)^n}.
$$
\end{definition}

\begin{theorem}\label{mainconv}
The $\dia$--product is well-defined and makes the space of convolutes
of $\CS'$ into an algebra over $\CC$.
\end{theorem}
\begin{proof} We will prove that the $\dia$--product of two convolutes of $\CS'$ is 
well-defined, and is again a convolute of $\CS'$. The rest is obvious.

It is sufficient to consider the case when
$$
{\widetilde f}(p)=D^\alpha {\widetilde f}_0(p),\quad {\widetilde g}(p)=D^\beta 
{\widetilde g}_0(p).
$$
Then, integrating by parts, we may rewrite the $\dia$--product in the following form:
$$
(-1)^{|\alpha|}\int\ {\widetilde f}_0(q)\ \frac{\partial^\alpha}{\partial 
q^\alpha}
\left[\frac{\partial^\beta}{\partial p^\beta}\ {\widetilde g}_0(p-q)\  
e^{\sqrt{-1}\omega(p,q)}\right]\ \frac{d^nq}{(2\pi)^n}.
$$
Derivatives acting on the exponential bring down powers of $p,$ so
the integral can be rewritten as
$$
P\left(p,\frac{\partial}{\partial p}\right)\int\ {\widetilde f}_0(q)\ 
\frac{\partial^\beta}{\partial p^\beta}\ {\widetilde g}_0(p-q)\ e^{\sqrt{-1}\omega(p,q)}
\ \frac{d^nq}{(2\pi)^n},
$$
where $P(u,v)$ is a homogeneous polynomial of degree $|\alpha|$. We now use the
Leibniz rule repeatedly to rewrite the expression above as
$$
P\left(p,\frac{\partial}{\partial p}\right)\int\ Q\left(q,
\frac{\partial}{\partial p}\right)\left[{\widetilde f}_0(q)\ 
{\widetilde g}_0(p-q)\ e^{\sqrt{-1}\omega(p,q)}\right]\ \frac{d^nq}{(2\pi)^n},
$$
where $Q(u,v)$ is a homogeneous polynomial of degree $|\beta|$. Because both 
${\widetilde f}_0$ and ${\widetilde g}_0$ are rapidly decreasing, the integral
converges absolutely and defines a $C^0$ function of $q$ which is rapidly
decreasing. Hence the $\dia$--product of ${\widetilde f}_0$ and ${\widetilde g}_0$
has the form
$$
\sum_{|m|\leq|\alpha|+|\beta|} D^m {\widetilde h}_m(p), 
$$
where the functions ${\widetilde h}_m(p)$ are continuous and rapidly decreasing.
It follows that the space of convolutes is closed under the $\dia$-product.
\end{proof}

\begin{cor} The space of multipliers of $\CS'$ inherits a product 
from the $\dia$--product on the space of convolutes of $\CS'$, and this product makes
the space of multipliers into an algebra over $\CC.$ Polynomials form
a subalgebra of this algebra isomorphic to the Weyl algebra with generators
$x_i,\ i=1,\ldots,n,$ and relations
$$[x_i,x_j]=2\sqrt{-1}\omega_{ij}.$$
\end{cor}
\begin{proof}
The first statement is an immediate consequence of theorems~\ref{isoconv} and~\ref{mainconv}.
The second statement follows from a simple computation.
\end{proof} 

It is this product on the space of multipliers that we call the Wigner-Moyal product
and denote with $\star$.

%
%
%
%
%
%
%
%
%
%
%
%
%
%

\section*{Acknowledgements}
We are grateful to A.~Beilinson, V.~Ginzburg, 
L.~Katzarkov, N.~Nekrasov, T.~Pantev, and A.~Yekutieli 
for useful discussions.
We also wish to thank the Institute for Advanced Study, Princeton, NJ,
for a very stimulating atmosphere.

\end{document}